\documentclass[twocolumn,twocolappendix]{aastex7}
\usepackage{amsmath}
\usepackage{booktabs}

\usepackage{tabularx}

\def\r#1{\textcolor{red}{#1}}
\newcommand{\wcen}{\ensuremath{\omega}\,Cen}

\newcommand{\mcomp}{4.46}
\newcommand{\e}{0.72}
\newcommand{\cosi}{0.43}
\newcommand{\avis}{31.1}
\newcommand{\Period}{94}
\newcommand{\longasc}{2.87}
\newcommand{\argperi}{1.60}
\newcommand{\Tperi}{2012.2}
\newcommand{\deltaRA}{0.24}
\newcommand{\deltaDec}{0.46}
\newcommand{\deltapmra}{-0.01}

\newcommand{\mvis}{0.78}
\received{May 4, 2026}
\revised{June 1, 2026}
\accepted{June 6, 2026}

\begin{document}

\title{A Long Period Stellar-Mass Black Hole Binary in $\omega$ Centauri}

\author[orcid=0009-0007-3015-9900,gname='Matthew',sname='Whitaker']{Matthew Whitaker}
\affiliation{Department of Physics and Astronomy, University of Utah, Salt Lake City, UT 84112, USA}
\email[show]{matthew.whitaker@utah.edu}  

\author[orcid=0009-0006-3217-8796]{Evan Kerr}
\affiliation{Department of Physics and Astronomy, University of Utah, Salt Lake City, UT 84112, USA}
\email{}

\author[orcid=0000-0003-0248-5470]{Anil Seth}
\affiliation{Department of Physics and Astronomy, University of Utah, Salt Lake City, UT 84112, USA}
\email{aseth@astro.utah.edu}

\author[orcid=0000-0002-5844-4443]{Maximilian Häberle}
\affiliation{European Southern Observatory}
\email{}

\author[orcid=0000-0002-1468-9668]{Jay Strader}
\affiliation{Michigan State University}
\email{}

\author[orcid=0000-0003-2861-3995]{Jay Anderson}
\affiliation{Space Telescope Science Institute}
\email{}

\author[orcid=0000-0003-3858-637X]{Andrea Bellini}
\affiliation{Space Telescope Science Institute}
\email{}

\author[orcid=0009-0005-8057-0031]{Callie Clontz}
\affiliation{Max-Planck-Institut f\:ur Astronomie}
\email{}

\author[orcid=0009-0002-5295-8504]{Zack Freeman}
\affiliation{Department of Physics and Astronomy, University of Utah, Salt Lake City, UT 84112, USA}
\email{}

\author[orcid=0000-0002-5060-1379]{Massimo Griggio}
\affiliation{Space Telescope Science Institute}
\email{}

\author[orcid=0000-0001-6604-0505]{Sebastian Kamann}
\affiliation{Liverpool John Moores University}
\email{}

\author[orcid=0000-0001-9673-7397]{Mattia Libralato}
\affiliation{INAF -- Padova}
\email{}

\author[orcid=0000-0002-6922-2598]{Nadine Neumayer}
\affiliation{Max-Planck-Institut f\:ur Astronomie}
\email{}

\author[orcid=0000-0002-0933-6438]{Elena Gonz\'alez Prieto}
\affil{Northwestern University}
\affil{Center for Interdisciplinary Exploration \& Research in Astrophysics (CIERA)}
\email{}

\author[orcid=0000-0003-4175-8881]{Carl L.~Rodriguez}
\email{}
\affil{University of North Carolina at Chapel Hill}

\author[orcid=0000-0003-4746-6003]{Sara Saracino}
\affil{Osservatorio Astrofisico di Arcetri}
\email{}

\author[orcid=0000-0002-7489-5244]{Peter Smith}
\affiliation{Max-Planck-Institut f\:ur Astronomie}
\email{}

\author[orcid=0000-0003-4546-7731]{Glenn van de Ven}
\affil{University of Vienna}
\email{}

\author[orcid=0000-0003-2512-6892]{Zixian Wang}
\affiliation{Department of Physics and Astronomy, University of Utah, Salt Lake City, UT 84112, USA}
\email{}

\begin{abstract}

Modern simulations of stellar dynamics in globular clusters peg a dominant role for stellar-mass black holes, but direct evidence for black holes in clusters remains limited. We present the discovery of an astrometric stellar-mass black hole--main sequence star binary in $\omega$ Centauri, the most massive Galactic globular cluster, using Hubble Space Telescope data from the oMEGACat project and additional JWST data that span a total of 23 years. The luminous companion to the black hole is a main-sequence turnoff star, and has a period of $\Period^{+63}_{-42}$~years, a semi-major axis of $31^{+15}_{-12}$~AU, and an eccentricity of $e$=$\e^{+0.08}_{-0.13}$. Since we observe the binary during periastron, the mass of the black hole is well-constrained even though we only observe a partial orbit: the inferred black hole mass is $\mcomp^{+1.22}_{-1.01}$~M$_\odot$.  We call this black hole oMEGACat BH-2. This is the first astrometric discovery of a stellar-mass black hole in a globular cluster, and is the longest period black hole binary system yet discovered. The low mass of this black hole is perhaps surprising given the low metallicity of the cluster, and shows that at least some low-mass black holes form at metallicity $Z<10^{-3}$. We find that the binary is almost certainly dynamically formed and is soft, with an expected binary disruption timescale of $\sim$800~Myr. While the total number of black hole binaries in $\omega$ Centauri is uncertain, we show that existing surveys only cover a small area of parameter space, and that the presence of additional detectable black hole binaries is likely.

\end{abstract}

\section{Introduction} \label{sec:introduction}

For at least 50 years, it has been evident that globular clusters contain an overabundance of X-ray sources formed through dynamical interactions  \citep{Clark1975,Katz1975,Fabian1975}. Subsequent observational work (e.g., \citealt{Hertz1983}) showed that luminous X-ray sources in Galactic globular clusters---both persistent and transient---are mostly or entirely accreting neutron stars.

Some of the massive stars formed early in the lives of star clusters should have left stellar-mass black holes upon their deaths rather than neutron stars. The black holes with small natal kicks will remain in the cluster and mass segregate within a fraction of the relaxation time of the cluster \citep{Merritt2013}, typically $<$1~Gyr. 
For decades it was predicted that these mass-segregated black holes would interact to form binaries and mutually eject each other, leading to few or no black holes remaining in current Galactic globular clusters (e.g., \citealt{Spitzer1969,Kulkarni1993, Sigurdsson1993}). 

Gradually it was realized that black holes do not fully decouple from the other stars, but instead interact efficiently with the surrounding cluster. This leads to less efficient black hole ejection than previously argued, and suggests the possibility of ongoing black hole populations in some clusters (e.g., \citealt{Mackey2008, Morscher2013, Sippel2013}). Parallel observational work identified candidate accreting black holes in both extragalactic and Galactic globular clusters (e.g., \citealt{Maccarone2007,Strader2012,Chomiuk2013,Miller-Jones2015}). These theoretical and observational efforts took on a broader significance with the gravitational wave discovery that black hole--black hole mergers are common \citep{Abbott2016} and the realization that many of these systems could have dynamically formed in massive star clusters (e.g., \citealt{Rodriguez2016}). Modern modeling of globular cluster structural properties or dynamics gives indirect evidence for black hole populations in some clusters (e.g., \citealt{Weatherford2020,Vitral2022,Dickson2024}).

The contemporary theoretical view of the dynamical evolution of globular clusters now puts black holes at the center, both literally and figuratively. Clusters with large cores are thought to have retained a substantial population of black holes, while traditional ``core collapse" is only possible once all (or nearly all) black holes have been ejected (e.g., \citealt{Breen2013,Heggie2014,Wang2016,Kremer2018,Kremer2019}). 

While the dynamics of black holes are relatively straightforward, the expected starting black hole population is uncertain, as it depends on several poorly constrained physical parameters. For example, the dependence of the initial mass function of massive stars on metallicity is poorly constrained (e.g., \citealt{Cameron2024}), as is the mapping between initial mass and remnant mass (e.g., \citealt{Boccioli2024}). Another uncertainty is the distribution of kick velocities; at least some black holes receive either natal or Blaauw kicks sufficiently large to remove them from globular clusters (see the recent review of \citealt{Popov2025}). 

Perhaps surprisingly, large variations in black hole populations are not necessarily predicted to translate into comparable differences in the number of observable black hole--visible star binaries in these clusters, with theoretical studies predicting weak (e.g., \citealt{Morscher2015}) or no (e.g., \citealt{Chatterjee2017}) correlation between these quantities. This is a consequence of the forecasted effects of black holes on the central dynamics: when a large black hole population is present, the central stellar density is expected to be low, such that black holes rarely interact with stars. But once most black holes have been ejected, the central stellar density increases and the few remaining black holes are much more likely to form binaries with normal stars.

Despite the broad implications of this shift in the theoretical view of cluster dynamics, there remains scant direct observational proof of the presence of black holes in globular clusters. Only two stellar-mass black holes have been dynamically confirmed in Galactic globular clusters: both in the metal-poor cluster NGC 3201, via an untargeted radial velocity survey \citep{Giesers2018,Giesers2019}. These are both detached binaries with main sequence secondaries, one in a 167 day, very eccentric orbit, and the other in a 2.2 day, low-eccentricity orbit. While the black hole masses are uncertain up to an inclination factor, they are not too large ($\lesssim 10 M_{\odot}$) unless the orbits are face-on. In addition to these systems, there is strong evidence that Gaia BH3, which hosts a $33 M_{\odot}$ black hole in a wide 11.6 yr orbit \citep{GaiaCollaboration2024}, is associated with a very metal-poor dissolved globular cluster tidal stream \citep{Balbinot2024} and hence could reasonably be included in a globular cluster black hole census.

Radial velocity studies with data comparable to that for NGC 3201 have been undertaken for 47 Tuc \citep{Muller-Horn2025} and $\omega$ Centauri \citep{Saracino2025}, but uncovered no candidate black hole binaries. These findings are intriguing given that, as discussed above, the occurrence of observable black hole binaries is predicted to be mostly independent of the present-day black hole population. This would imply similar numbers of black hole--visible star binaries should be found in most massive clusters. 

However, these radial velocity surveys are incomplete: they are insensitive to longer orbital periods and miss fainter stars (in both clusters the 50\% completeness of the binary search was around the main sequence turnoff). Since models for dynamical formation of black hole--visible star binaries predict a wide range of secondary masses \citep{Kremer2018}, surveys that cannot reach typical cluster main sequence stars are expected to miss most black hole--non black hole binaries.

Astrometric searches for black hole--non black hole binaries are complementary to radial velocity searches. Hubble Space Telescope (HST) or JWST data typically have sensitivity to much fainter and lower mass stars than ground-based radial velocities. In addition, astrometric searches are more sensitive to longer periods.

The highest-quality space-based astrometric dataset for any Galactic globular cluster is for $\omega$ Centauri ($\omega$ Cen): \citet{Haberle2024} present hundreds of epochs of HST astrometric measurements over a typical baseline of more than 20 years. This remarkable dataset has already been used to demonstrate the existence of an intermediate-mass black hole at the center of $\omega$ Cen \citep{Haberle2024IMBH}. These same data are also ideal to search for astrometric binaries in the cluster.

Fortuitously, $\omega$~Cen also likely hosts the largest population of stellar mass black holes of any cluster in the Milky Way. This is for three reasons: first, it is the most massive cluster, so plausibly had the highest initial black hole population; second, the high escape velocity \footnote{The current escape velocity is 62 km s$^{-1}$ \citep{Haberle2024IMBH}, compared to a median of 28 km s$^{-1}$ for Galactic globular clusters \citep{Gnedin2002}. However, the $\omega$ Cen escape velocity was potentially $\gtrsim 100$ km s$^{-1}$ when the black holes formed, as clusters lose substantial mass due to stellar evolution, two-body relaxation, and shocks (e.g., \citealt{Fall2001}).} enables retention of black holes after natal or merging kicks, and third, the cluster has a long core relaxation time (a few Gyr, \citealt{vandeVen2006}), slowing dynamical ejection of black holes.

Indeed, dynamical modeling of cluster kinematics and pulsar accelerations in $\omega$~Cen suggest the presence of $\mathcal{O}(10^4)$ black holes \citep{Zocchi2019, Baumgardt2019, Dickson2024, Banares-Hernandez2025}. Since these papers did not self-consistently
include an intermediate-mass black hole of the mass inferred by \citet{Haberle2024IMBH} together with a separate population of stellar-mass black holes, this population approximation is likely an overestimate. Nonetheless, for the reasons given above, the black hole population is still expected to be large.

\added{$\omega$~Cen's  large spread in metallicity \citep{Norris1996,Johnson2010} and age \citep{Clontz2024}, and the detection of a central intermediate-mass black hole \citep{Haberle2024}, make a strong case that it is a stripped nuclear star cluster from an early merger in the Milky Way's history \citep[e.g.][]{Bekki2003,Pfeffer2021,Limberg2022}.  The deep potential wells of nuclear star clusters, as well as the presence of central massive black holes, can impact the dynamical evolution of stellar mass black holes in these systems \citep[e.g.][]{Merritt2009,Antonini2016, Fragione2019,Panamarev2019}. In addition, because nuclear star clusters can assemble via repeated cluster accretion or star formation events \citep[see review by][]{Neumayer2020}, this assembly history may impact the compact object population in the cluster. For example, progenitor star clusters with shorter relaxation times could mass segregate before falling into the nuclear star cluster, leading to a more dynamically evolved compact object population in the nuclear cluster than would otherwise be expected based solely on its present mass and radius (e.g., \citealt{Rantala2024,Bernadich2026}).}

Despite the likely presence of many stellar mass black holes in $\omega$~Cen, none have been detected so far. This is despite deep radio \citep{Tremou2018,Mahida2025} and X-ray surveys \citep{Haggard2013} that would be sensitive to accreting black holes \citep[e.g.][]{Strader2012,Chomiuk2013}, as well as radial velocity and astrometric searches for binary stars with black hole companions \citep{Wragg2024,Platais2024, Saracino2025}.

\citet{Platais2024} used an independent reduction of a subset of the long-baseline $\omega$~Cen HST astrometric dataset to identify four candidate astrometric binaries with ``dark" companions, which they inferred to be white dwarfs or possibly neutron stars. 

We have undertaken a new search for astrometric binaries in $\omega$~Cen using the more extensive \citet{Haberle2024} HST dataset, augmented with new JWST data. The binaries found from this effort will be presented in an upcoming paper (Whitaker et al., {\em in prep}). One of the highest significance binaries found in our search is a recovery of one of the ``dark" companion candidates identified by \citet{Platais2024}, who suggested a likely neutron star companion. However, here we show that it is in fact a stellar-mass black hole--main sequence star binary with a long ($94^{+63}_{-42}$ yr) orbital period.

Section \ref{sec:data} describes the data and simulations used in this discovery and its interpretation. Section \ref{sec:starprops} contains a detailed analysis of the properties of the main sequence star, which are needed to interpret the orbital properties of the binary. Section \ref{sec:results} describes our constraints on the binary properties. In Section \ref{sec:discussion}, we explore the implications of this discovery for the black hole population in $\omega$ Cen and in other massive clusters.  Throughout the paper, we use a distance to $\omega$~Cen of 5494$\pm$61~pc \citep{Haberle2025}.  At this distance, 1 mas/yr translates to 26.06 km s$^{-1}$ and 1 mas is 5.494 AU.  

\section{Data and Simulations} \label{sec:data}

This section describes the data used in the astrometric discovery of a long period black hole--main sequence binary in $\omega$ Cen. The evidence for this classification is presented in Section \ref{sec:results}.

\subsection{Space-Based Astrometry and Photometry}
\label{subsec:astrophotometry}
\subsubsection{HST Astrophotometric Data}
Here we briefly describe the reduction of the HST images, and refer readers to \citet{Haberle2024} for a detailed description.

The individual \added{charge transfer efficiency corrected \texttt{*\_flc.fits}} images are first processed with \texttt{hst1pass} \citep{Anderson2022} which refines library effective PSF (ePSF) models for each image. After applying geometric distortion corrections, this gives an initial catalog of sources for each image. These single-image catalogs are then crossmatched with \textit{Gaia} EDR3 \citep{GaiaCollaboration2021,Lindegren2021} to determine the transformations that bring each image to a \textit{Gaia}-based astrometric master frame. Then, the code \texttt{KS2}  \citep[see][]{Bellini2017I} is used to perform PSF photometry using the refined ePSF from \texttt{hst1pass}, including source detection in combined images and iterative neighbor subtraction. KS2 yields astrometric measurements in the raw detector frame of each individual image. These are iteratively transformed to the astrometric master frame during the proper motion determination, yielding a position in the master frame in pixel coordinates and a photometric measurement for each observed star at each single image epoch. 

\begin{figure*}[ht!]
    \centering
    \includegraphics[width=1\linewidth]{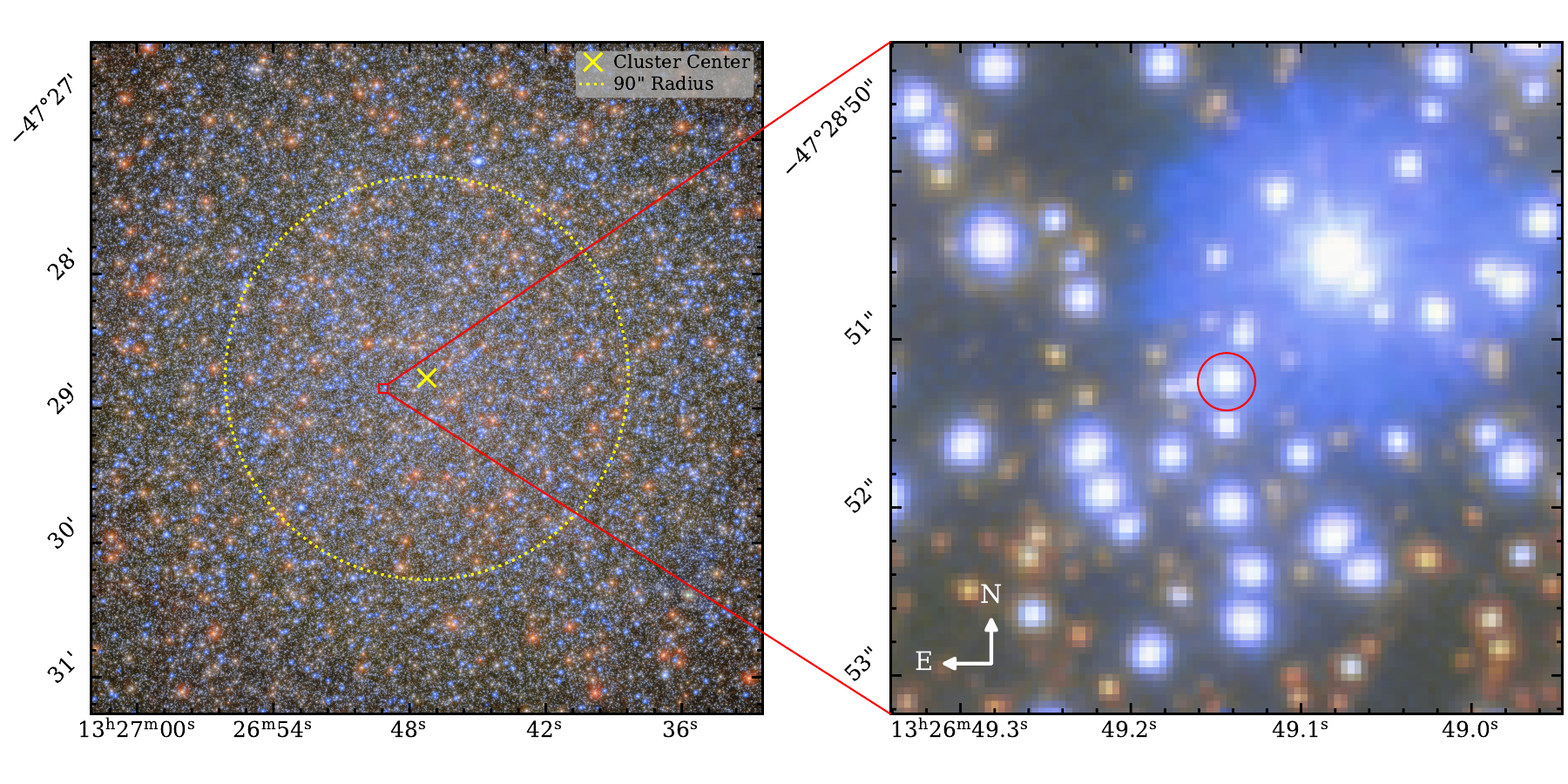}
    \caption{Stacked HST composite image of $\omega$~Cen. {\em Left} -- Location of the visible star relative to the cluster center \citep{Anderson2010, Haberle2024IMBH} {\em Right} -- A closer view of the visible star and its neighbors, including a bright horizontal branch star $\sim$1\arcsec\ to the northwest. \textit{Image Credit: ESA/Hubble, NASA, Maximilian H{\"a}berle (MPIA)}}
    \label{fig:star_hst_cutout}
\end{figure*}

For this work, we utilize the same epochs used for astrometry by \citet{Haberle2024}; that paper did not include individual exposure astrometry, instead combining astrometry into a single proper motion catalog.  Of the 672 possible astrometric measurements, we cull out data with (1) short (5s) exposure times, (2) those in which our black hole binary candidate (ID 285597 in the \citealt{Haberle2024} catalog) is not on the detector or not detected, (3) those where diffraction spikes from bright stars are present, and (4) those impacted by persistent noisy pixels. In total, we are left with 351 HST epochs spanning 20.5968 years. We also use an updated, empirical HST astrometric uncertainty model derived from repeat measurements that depends both on instrumental magnitude and accounts for the impact of crowding. The quality cuts and astrometric uncertainty model are further detailed in our upcoming work (Whitaker et al., {\em in prep}).  Fig.~\ref{fig:star_hst_cutout} shows multi-band HST imaging of our black hole binary candidate.  It lies in the crowded core of the cluster, but is brighter than any other star within 0$\farcs$5. Table \ref{tab:astrometry} contains the HST astrometric data used. \added{The HST data is available from the Mikulski Archive for Space Telescopes (MAST) at the Space Telescope Science Institute. The utilized observations can be accessed via \dataset[doi: 10.17909/26qj-g090]{https://doi.org/10.17909/26qj-g090}.}

\subsubsection{JWST Astrometry}

To augment the existing HST astrometry for our candidate, we use two epochs of JWST NIRCam Short Wavelength Channel observations. The star is present in the F200W images at epoch 2024.5219 (GO 4343, \citealt{2023jwst.prop.4343K, 2026ApJ..1001..127C}; 4 dithers totaling 515s of exposure time). It is also present in F115W and F200W images at epoch 2025.5777 in the astrometric monitoring program of the center of $\omega$~Cen (GO 8322, \citealt{2025jwst.prop.8322H}; 4 dithers of 419s exposures in each filter).  We do not use the existing long wavelength channel observations due to their lower spatial resolution, which leads to decreased astrometric precision.
The details of the JWST data reduction will be described in a future publication (H{\"a}berle et al., \textit{in prep}.), but overall follow the same strategy as for the HST observations. The JWST data are processed using the JWST pipeline up to Stage 2 (\texttt{*.cal.fits}), i.e. they are flat-fielded and flux-calibrated, but not remapped. An initial photometry pass is performed on the individual images using \texttt{jwst1pass} (see \citealt{Anderson2022,2023ApJ...950..101L}) and public library geometric distortion corrections and empirical point-spread-function models\footnote{\url{https://www.stsci.edu/~jayander/JWST1PASS/LIB/}}. We then use a modified version of \texttt{KS2} to perform our final astro-photometric measurements.\\
To cross-match and transform the new JWST positions to the same reference frame as the \cite{Haberle2024} HST data, we use the same set of well-measured cluster-member stars, propagated to the respective JWST epoch. For each star, we determine the closest 50 of these reference stars and determine the best fit linear six-parameter transformations to transform detector coordinates to reference frame coordinates. This yields a total of 12 individual astrometric measurements for the black hole binary candidate.

To calculate the uncertainties for each measurement, we measure the standard deviation (with $N-1$ degrees of freedom) of the positions from the four images in each epoch and filter. These standard deviations are between 0.129 and 0.777 mas. The more precise measurements are from the 2024 data, while the 2025 data have larger scatter due to a different detector readout strategy, for which this fairly bright star was only unsaturated in the "Frame 0" read.
The individual JWST astrometric measurements can be found at the end of Table \ref{tab:astrometry}. \added{The specific JWST observations analyzed can be accessed from MAST via \dataset[doi: 10.17909/32xm-hy79]{https://doi.org/10.17909/32xm-hy79}.}

\begin{table*}
\caption{HST and JWST astrometric measurements of the visible star. Measurements are relative to the \citet{Haberle2024} catalog position and do not include the motion of {\wcen}. Entries with a non-zero flag are not used in the orbital fit. Flag descriptions: 1: Large ($>5\sigma$) outlier; 2: Overlapping diffraction spike; 3: Centroid falls on noisy pixel; 4: Short ($<5$s) exposure time.}
\begin{tabular}{cccccccc}
\hline
Year & $\Delta$ RA [mas] & $\Delta$ Dec [mas] & $\Delta$ RA err [mas] & $\Delta$ Dec err [mas] & Instr. & Filter & Flag \\
\hline
\hline
2002.48916 & -0.832 & -0.296 & 1.014 & 1.014 & ACS & F625W & -- \\
2009.53802 & -1.472 & -2.496 & 1.430 & 1.430 & UVIS & F336W & -- \\
2010.03538 & -0.352 & -0.496 & 0.453 & 0.453 & UVIS & F606W & -- \\
2010.0403 & 0.048 & -1.216 & 0.506 & 0.506 & UVIS & F336W & -- \\
2010.3243 & -0.472 & -1.096 & 0.489 & 0.489 & UVIS & F555W & -- \\
2010.3272 & -0.472 & -0.976 & 0.548 & 0.548 & UVIS & F814W & -- \\
2010.50723 & 1.168 & -0.776 & 0.510 & 0.510 & UVIS & F336W & -- \\
2011.12553 & -0.552 & -0.336 & 0.453 & 0.453 & UVIS & F606W & 2 \\
2011.21868 & -0.032 & -0.456 & 0.454 & 0.454 & UVIS & F606W & 2 \\
2012.18571 & -0.952 & -0.456 & 0.456 & 0.456 & UVIS & F606W & -- \\
2012.6314 & -1.152 & 1.304 & 0.383 & 0.383 & ACS & F435W & -- \\
2013.95214 & -0.432 & 1.744 & 0.433 & 0.433 & ACS & F814W & -- \\
2015.43183 & 0.768 & 0.464 & 0.454 & 0.454 & UVIS & F606W & -- \\
2016.20181 & 0.928 & 3.144 & 0.456 & 0.456 & UVIS & F606W & 2 \\
2019.16645 & 2.808 & 2.464 & 0.386 & 0.386 & UVIS & F814W & -- \\
2021.00775 & 4.648 & 2.984 & 1.095 & 1.095 & UVIS & F606W & 4 \\
2022.03218 & 4.168 & 1.904 & 0.424 & 0.424 & UVIS & F606W & -- \\
2025.5777 & 6.300 & 1.332 & 0.596 & 0.777 & NIRCAM & F115W & -- \\
\hline
\end{tabular}\\

\textbf{Note:} A representative sample of observations are shown here. A complete machine-readable table will be made available.
\label{tab:astrometry}
\end{table*}

\subsection{MUSE Spectroscopy}

The visible star has 31 3x15s MUSE exposures taken on 12 distinct dates ranging from 2015-04-23 to 2022-05-31. These data have already been analyzed and described in \citet{Nitschai2023} (MUSE Star ID: 4420) and \citet{Saracino2025}, and we refer readers to those papers for reduction details. Spectra were obtained using the PSF fitting software PampelMUSE \citep{Kamann2013}, which enables extraction of stellar spectra  in crowded fields, especially when the positions of those stars are well known from HST imaging.

\subsection{Simulation Outputs}
\label{sec:simulation}

To provide context for the black hole binary, we use outputs from recent dynamical simulations of an $\omega$ Cen-like cluster by \citet{GonzalezPrieto2025}.  These Cluster Monte Carlo simulations include an intermediate-mass black hole and have been fit to $\omega$~Cen's present day properties.  We analyze the simulation that started with a 500~M$_\odot$ central BH seed here; at the end of the simulation this seed has grown to an intermediate-mass black hole with a mass of 4.7$\times$10$^4$~M$_\odot$ and the cluster contains 2459 stellar mass black holes. The simulation tracks the formation and destruction of binaries, including those involving black holes. The simulation includes 9 visible star--black hole binaries at the ending timestamp of the simulation (12.5~Gyr) with black hole companions ranging from 9 to 27~M$_\odot$. To further increase number statistics on the simulated population of visible star--black hole binaries, we also analyzed 50 snapshots taken every 20~Myr over the last Gyr of the simulation, yielding 474 measurements of visible star--black hole binaries.  Available data on 
these simulated binaries includes their star and black hole masses, the orbital period and semi-major axis of the binary system, and their position and velocity within the cluster. 

We use the one-dimensional velocity dispersion of all 474 visible star--black hole binaries (12.6 km s$^{-1}$) to set a prior on the proper motion of the system (Sec.~\ref{sec:orbital_fit}); this value is lower than the velocity dispersion observed for the stars due to partial energy equipartition.  We also use the simulations to gauge the expected lifetime of the detected binary (Section~\ref{sec:fewbody-simulations}), and to understand whether binary systems like the one we detect here are consistent with predictions from these simulations (Section \ref{sec:wcen_bh_population}).  

\subsection{Multi-wavelength Constraints}

We show in this paper that star 285597 from \citet{Haberle2024} is in a long period binary with a stellar-mass black hole. Given the long orbital period and that the secondary is a $\sim 0.8 M_{\odot}$ main sequence star, no accretion onto the black hole is expected. Nonetheless, for completeness, here we consider multi-wavelength constraints on accretion.

\subsubsection{X-ray}
\label{sec:X-ray}

The binary is in the region of the cluster covered by the Chandra/ACIS-I survey of \citet{Henleywillis2018}, which reached an unabsorbed limiting depth of $\sim 2 \times 10^{-16}$ erg s$^{-1}$ cm$^{-2}$ (0.5--4.5 keV) near the cluster center. There is no X-ray source within $\sim 30\arcsec$ of the binary. Assuming a typical power-law index photon flux ($N_E \propto E^{-\Gamma}$ with $\Gamma = 1.8$), the non-detection flux limit is equivalent to a 1--10 keV luminosity limit of $\lesssim 9 \times 10^{29}$ erg s$^{-1}$ at the distance of $\omega$ Cen. This upper limit is comparable to the faintest black hole X-ray binaries known (e.g., \citealt{Rodriguez2020})  and is $>$2 orders of magnitude below the candidate black hole X-ray binaries identified in M62 \citep{Chomiuk2013} and 47 Tuc \citep{Miller-Jones2015}, though as stated above no accretion from the visible companion is expected.

\subsubsection{Radio}

The region containing the binary is covered by a number of existing deep cm radio continuum images of the cluster. It is not detected in 29.4 hr of on-source imaging with the Australia Telescope Compact Array at a central frequency of 7.25 GHz, giving a $3\sigma$ upper flux density limit $<8.8 \mu$Jy \citep{Tremou2018,Tudor2022}. This flux density limit is equivalent to radio luminosity $L_R \lesssim 2.3 \times 10^{27}$ erg s$^{-1}$ at 7.25 GHz, or $L_R \lesssim 1.6 \times 10^{27}$ erg s$^{-1}$ at the standard comparison frequency of 5 GHz, if a flat radio spectrum between 7.25 and 5 GHz is assumed. 
The recent deeper Australia Telescope Compact Array 7.25 GHz image of \citet{Mahida2025} also does not appear to detect a source at the location of the binary (see their Figure 1). Given their $3\sigma$ limit of $<3.3\mu$Jy, this would lower these upper limits by a factor of about 2.7, to a 5 GHz limit of $L_R \lesssim 6.0 \times 10^{26}$ erg s$^{-1}$. As for the X-ray data, these limits are deep, within a factor of a 
few of the very low-level radio emission detected from quiescent stellar-mass black holes such as A0620-00 \citep{Gallo2006,Dincer2018}.

\section{Results: Properties of the Visible Star}
\label{sec:starprops}

To determine the properties of the luminous star in the binary, we use existing oMEGACat data, including both multi-band photometry and astrometry from \citet{Haberle2024} and MUSE spectroscopic analysis from \citet{Nitschai2023}. As we show below (Section \ref{sec:dynamical-evolution}), the binary is likely dynamically formed, thus the visible star's properties are important primarily to constrain the stars mass in the binary orbital fit.

\subsection{HST Astrophotometric Measurements} 

From the catalog of \citet{Haberle2024}, the visible star is located at ICRS (R.A., Dec.) of ($13^{\mathrm{h}}26^{\mathrm{m}}49\fs1457528$,$-47^\circ28'51\farcs231396$) at an Epoch of 2002.5\footnote{Due to setup of the reference frame which is co-moving with Omega Centauri, the reference epoch for absolute positions is 2002.5, while for position changes due to internal-motions it is 2012.0; see \cite{Haberle2024} for details.}, 20\arcsec\ (0.5~pc) southeast of the cluster center defined by \citet{Anderson2010} (Fig. \ref{fig:star_hst_cutout}). We note this cluster center is also consistent with the position of the intermediate-mass black hole \citep{Haberle2024IMBH}. The visible star's HST photometry \citep{Haberle2024} indicates it is near the main-sequence turnoff, with a WFC3/UVIS $F814W$  magnitude of 17.557$\pm$0.006; assuming a reddening of E(B-V)$=0.185$ \citep{Clontz2024}, the extinction corrected magnitude is $F814W = 17.214$. Its position in a $F275W-F814W$ vs.~$F814W$ color-magnitude diagram (CMD) is shown in the left panel of Fig.~\ref{fig:star_spectra_cmd}; the star occupies a comparable relative position in other CMDs we examined with the available seven-band photometry. 

The star is a proper motion member of the cluster, and the catalog proper motion implies a two-dimensional velocity of 11~km s$^{-1}$ relative to the cluster mean. However, as we discuss below, this proper motion measurement is partially degenerate with the binary motion of the visible star. For our best fit, we find a slightly lower total proper motion of 7~km s$^{-1}$ (Section \ref{sec:bestfit}).

Time series photometry shows no significant variations: the star does not show any evidence of being a variable star. 182 measurements are available in the WFC3/UVIS F606W filter, and these show no significant brightening or dimming over 13 years; colors are also constant between epochs. The star is several PSF FWHM ($0\farcs2$) away from the nearest star and is brighter than all stars within $0\farcs9$ (Fig.~\ref{fig:star_hst_cutout}), so uncontaminated photometric measurements of its properties are expected. 

Fig.~\ref{fig:star_hst_cutout} also shows that there is a luminous hot horizontal branch star located $\sim$1\arcsec\ northwest of the star (\citealt{Haberle2024} Star ID: 390029). While this star has no meaningful effect on the photometry \added{or astrometry\footnote{All neighboring stars, including this horizontal branch star, contribute minimal contaminating light ($<6\%$ across all wavelengths even without PSF modeling). We also see no filter-dependent offsets in the astrometry.}}, it does complicate our ground-based spectroscopy of the target star, as discussed below in Section \ref{sec:MUSEvis}. For this reason, a metallicity determined from stars which are photometrically similar to the visible star may be more accurate than the single measurement for the star itself. To determine a robust metallicity, we select cluster member stars within 0.02 magnitudes of the CMD position of the visible star that also have MUSE spectroscopic metallicities, giving a sample of 260 similar stars (henceforth we call these ``doppelgangers"). This sample of photometric doppelgangers has a mean [M/H] = $-1.75\pm0.02$ and $\sigma=0.25$ \citep[all metallicity values quoted here have atomic diffusion corrections applied;][]{Nitschai2023}.

The ``chromosome map" made from HST photometry can also provide some information on the helium content of the visible star \citep{Clontz2025}. The star's position on this diagram compared to other stars of similar brightness suggests it is a member of either the P1 (primordial) or Im (intermediate) cluster populations, consistent with either no or modest helium enrichment, and excluding high helium enrichment.

\subsection{MUSE Spectroscopic Measurements} 
\label{sec:MUSEvis}

\citet{Nitschai2023} analyzed an S/N-weighted mean spectrum of all available MUSE exposures for the visible star, finding a radial velocity of 237.1$\pm$4.2~km s$^{-1}$. This value is close to the systemic velocity of the cluster ($V_{sys} = 233$ km s$^{-1}$; \citealt{Nitschai2023}), independently confirming its cluster membership.

\begin{figure*}[ht!]    
    \centering
    \includegraphics[width=0.45\linewidth]{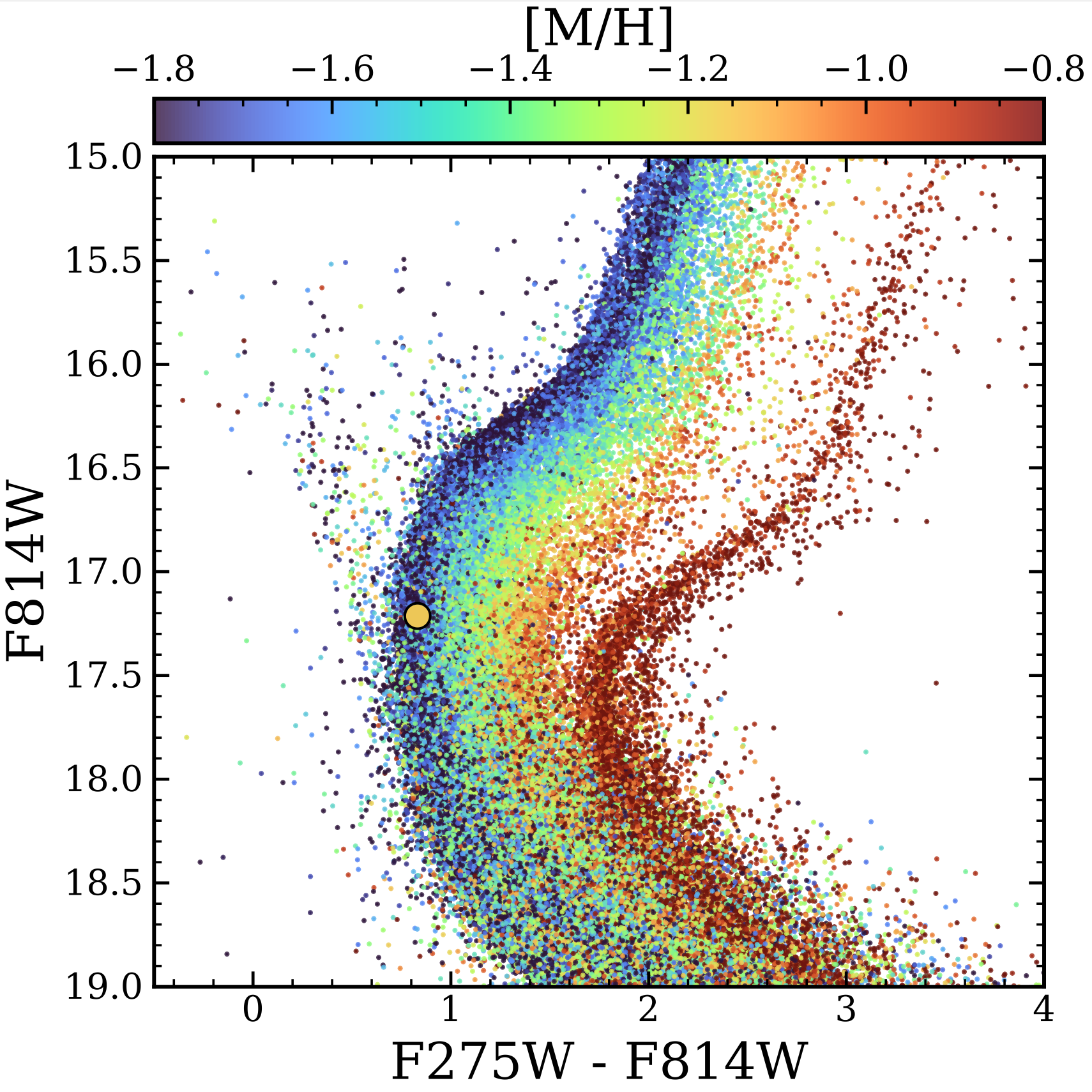}
    \includegraphics[width=0.43\linewidth]{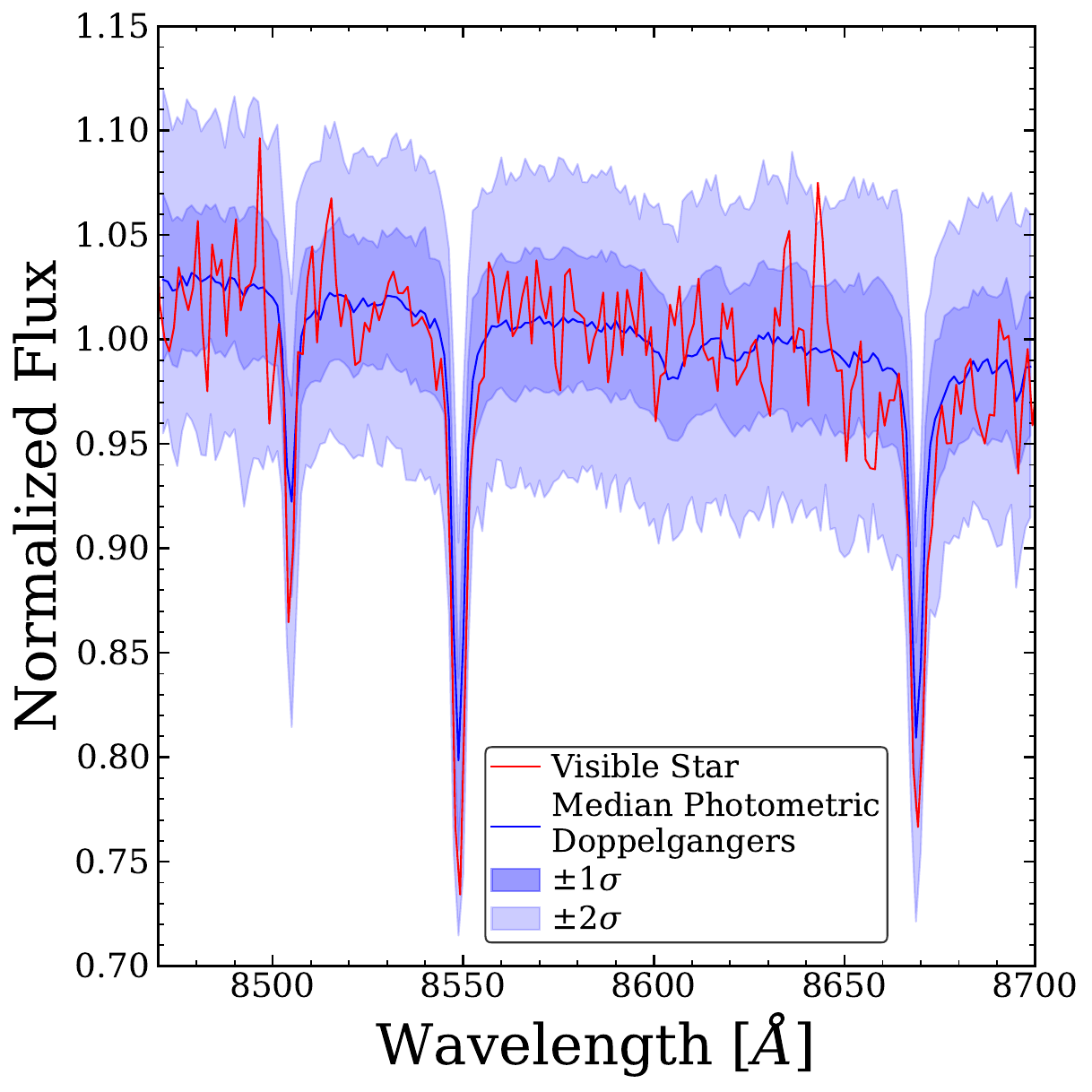}

    \caption{Visible star properties.  {\em Left --} Extinction corrected color-magnitude diagram combining photometry from \citet{Haberle2024} and spectroscopic metallicities from \citet{Nitschai2023}.  Based on its position in the CMD, it is a turnoff star with [M/H] = --1.75$\pm$0.25.   {\em Right} -- MUSE spectra of the visible star (red) vs. other stars with similar color (blue).  The blue line and shaded regions show the median, one, and two $\sigma$ range of the 260 stellar spectra selected 
    as photometric doppelgangers (within 0.02 magnitudes in both color and magnitude in the left CMD). The red line shows the calcium triplet lines are slightly deeper than for stars of similar color, and most consistent with [M/H] = --1.5. The combined photometric and spectroscopic data suggest [M/H] in the range --1.75 to --1.5.}
    \label{fig:star_spectra_cmd}
\end{figure*}

However, there is an apparent mismatch in its metallicity; the spectroscopic [M/H] derived for the star is $-1.15\pm0.09$, noticeably higher than the photometric doppelgangers
(Fig.~\ref{fig:star_spectra_cmd}).

To investigate this, we first compared the MUSE spectrum to that of the photometric doppelgangers, finding that the visible star's 
MUSE spectrum is much bluer. Visual inspection of the MUSE data cubes shows that at blue wavelengths, there is an overlap between the PSFs of the visible star and of the nearby hot horizontal branch star mentioned above. This is only an issue in the blue; the stars are well-separated at redder wavelengths. Inspection of the spectra extracted from individual MUSE exposures of the visible star shows a wide variation in the continuum slopes blueward of $\sim 7000 \, {\rm \mathring{A}}$, further suggesting the possibility of contamination by the horizontal branch star at bluer wavelengths in some of the data.

We therefore focus on the reddest end of the MUSE spectra, in the region of the calcium triplet ($\lambda \sim 8550{\rm \mathring{A}}$). We can confirm that contamination from the horizontal branch star is negligible at these wavelengths, as the hot star shows strong Paschen absorption that is not detected in the spectra of the visible star. A comparison of the calcium triplet region between the visible star and the photometric doppelgangers is shown in the right panel of Fig.~\ref{fig:star_spectra_cmd}. We find that the visible star has slightly deeper calcium triplet lines than the doppelgangers, suggesting a cooler and/or more metal-rich star. Compared to MUSE spectra of stars at similar magnitudes, we find the best match with stars that are redder by $\sim 0.2$ mag in $F275W-F814W$ and have [M/H] $\sim -1.5$. This value is consistent with the scatter of the metallicities of the doppelgangers ([M/H] = $-1.75$; $\sigma=0.25$) but inconsistent with the much higher spectroscopic metallicity of $-1.15\pm0.09$ in \citet{Nitschai2023}; this high value may perhaps be due to the neighbor contamination at bluer wavelengths. We therefore suggest that the spectroscopic catalog value is likely incorrect and that the true value is somewhere in the range $\sim$ --1.75 (the photometric estimate above calibrated from spectroscopy) to --1.5 (the calcium triplet spectroscopy estimate here). For context, the median metallicity of the cluster is [M/H] = $-1.61$ \citep{Nitschai2024}, thus the star is likely consistent with the cluster's dominant metallicity population.

Finally, we note that \citet{Saracino2025} derived radial velocities for the visible star in five separate epochs from the MUSE data.
While these measurements show no significant changes from the mean value presented in \citet{Nitschai2023}, they could potentially be affected by the contaminating hot star in the blue. 
Hence, we have re-derived velocities from these data using a two-step process. We first created a template in the calcium triplet region iteratively, starting from the  highest S/N exposure, and then using cross-correlation to find the best fit velocity offset for each spectrum, putting all at a common velocity. A weighted mean of these offset spectra was used to make the final template. We then used this template to get the relative velocity of the star in each individual exposure. We find 17 velocities over 12 nights with $S/N > 5$, and summarize these measurements in Table~\ref{tab:vrad}. A linear model fit to the velocities as a function of time finds a slope that is insignificant ($<1\sigma$), so there is no evidence for a linear trend in the velocities. Given the large uncertainties on the measurements, we do not incorporate these data directly into our orbital fits in Section \ref{sec:results}, but do use them as consistency checks in Section \ref{sec:improved_mass_constraints}.  

\begin{table}
\caption{MUSE radial velocity measurements of visible star}
\begin{tabular}{cccc}
\hline
Year & $V_r$ [km s$^{-1}$] & $V_r$ err [km s$^{-1}$] & S/N\\
\hline
\hline
2015.3075 & 1.9 & 5.8 & 16.1 \\
2015.3075 & --6.2 & 6.2 & 5.8 \\
2016.0909 & --4.4 & 12.7 & 5.2 \\
2016.0938 & 5.4 & 7.3 & 6.5 \\
2017.0749 & --0.2 & 9.2 & 11.3 \\
2017.0750 & 8.7 & 6.9 & 12.7 \\
2017.0832 & --2.7 & 6.5 & 8.0 \\
2017.0832 & --6.0 & 7.6 & 6.7 \\
2017.3123 & --2.4 & 7.5 & 8.9 \\
2017.3124 & 2.7 & 7.7 & 13.0 \\
2018.1214 & 6.4 & 11.4 & 5.7 \\
2018.2799 & 0.5 & 9.0 & 5.6 \\
2018.3593 & 4.8 & 6.0 & 5.3 \\
2019.1764 & --14.6 & 13.7 & 5.2 \\
2022.4083 & 8.5 & 13.8 & 8.6 \\
2022.4083 & 4.7 & 9.2 & 9.3 \\
2022.4113 & --1.7 & 11.3 & 7.1 \\
\hline
\end{tabular}\\

\textbf{Note:} All measurements are relative to a template with a velocity of 235.7$\pm$1.8 km s$^{-1}$; this should be added to the $V_r$ value to get the barycentric velocity.  
\label{tab:vrad}
\end{table}

\subsection{Visible Star Mass Estimate} 
\label{sec:star_est}
Based on the information above, we can infer the mass of the visible star. Following \citet{Haberle2025}, we use Dartmouth isochrones \citep{Dotter2008} to estimate stellar masses. For our fiducial mass estimate, we assume an age of 12 Gyr, the median age of the cluster \citep{Clontz2024}, a [Fe/H] of --1.7 and an [$\alpha$/Fe] of 0.3 \citep{Johnson2010} (corresponding to [M/H] $\sim -1.5$) and $Y = 0.25$ (a normal helium abundance for a metal-poor star). Using the F814W extinction corrected magnitude of 17.214, this isochrone yields a mass for the visible star of 0.78~M$_\odot$. The uncertain age and metallicity of the star create a small uncertainty on the stellar mass, but a much larger uncertainty comes from the uncertainty in the helium abundance. 

Stars in $\omega$~Cen can be enriched in helium \citep[e.g.][]{Piotto2005}, with a substantial population of stars having $Y \sim 0.4$ (often referred to as ``2G" or ``P2'' stars). At the metallicity of the visible star, roughly half of stars have these high helium values \citep{Clontz2025}. As noted above, the chromosome map for this star suggests it is unlikely to be strongly helium-enhanced. However, for completeness, we also derive the mass under the assumption that the star is part of the helium-enriched population.
Using an isochrone with identical parameters to that above, but with $Y=0.4$, we find a mass of 0.60~M$_\odot$, $\sim$0.18~M$_\odot$ lower than the inferred mass for a star with $Y=0.25$. In Section~\ref{sec:mass_constraints} we consider the impact on the mass of the black hole if the star is helium-enriched.

\section{Results: Orbital Fitting} \label{sec:results}

\begin{figure}[t!]
    \centering
    \includegraphics[width=1\linewidth]{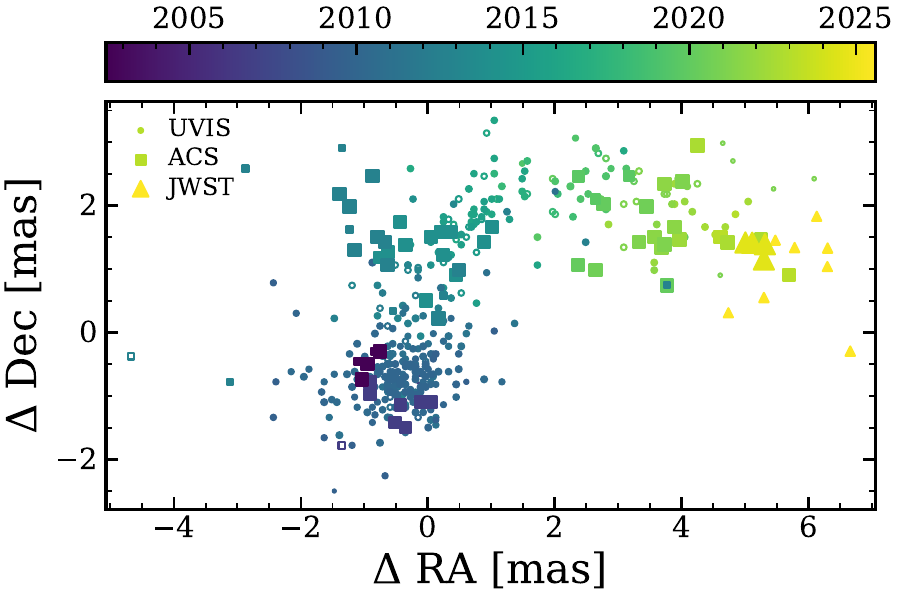}
    \caption{Visible star astrometry. Stellar positions measured by HST ACS/WFC (squares), WFC3/UVIS (circles) and JWST NIRCam (triangles). Color indicates the date of observation. The size of each marker is inversely proportional to the measurement's uncertainty. Open markers indicate epochs which were removed due to short ($<5$s) exposure time, potential contamination from overlapping diffraction spikes from a neighboring bright star, or the PSF falling on a more noisy pixel (Sec. \ref{subsec:astrophotometry}).}
    \label{fig:on_sky_data}
\end{figure}

 Using the combined astrometric data from HST and JWST, we see that our star (ID 285597) has a clear non-linear motion across the sky (Fig. \ref{fig:on_sky_data}). The strong acceleration of this source previously detected by \citet{Platais2024} is strengthened by the inclusion of the earlier ACS data, additional non-$F606W$ images, and the new epochs of JWST data. The observed motion is a combination of the proper motion of the binary and the orbit of the visible star around the binary center of mass. 

\subsection{Orbital Fits} \label{sec:orbital_fit}

 To determine the orbital characteristics of this system, we used Markov Chain Monte Carlo (MCMC) techniques to fit an 11 parameter orbital model to the astrometric data. Our model includes 7 parameters for the binary:
 \begin{itemize}
     \item the orbital period ($P$)
     \item semi-major axis of the visible star ($a_\text{vis}$)
     \item cosine of the inclination angle ($\cos i$)
     \item eccentricity ($e$)
     \item longitude of the ascending node ($\Omega$)
     \item argument of periastron ($\omega$)
     \item time of periastron passage ($T_0$)
 \end{itemize}
In addition, there are four parameters accounting for the proper motions relative to the mean motion of the cluster and astrometric offsets relative to the \citet{Haberle2024} catalog position. The catalog proper motions are $\mu_\text{cat,RA} = -0.38886$\,mas/yr and $\mu_\text{cat,Dec} = 0.28154$\,mas/yr. Because each measurement is made relative to the cluster itself, we expect no significant parallax signal for cluster member stars.

Our model uses the Thiele-Innes elements \citep{Lucy2014}, which are transformations of $a_\text{vis}$, $i$, $\Omega$, and $\omega$, to calculate the apparent ellipse of the luminous companion around the barycenter. We include the proper motions and astrometric offsets through a linear model for the right ascension and declination. The astrometric offset term accounts for small corrections to the catalog position and for the fact that unresolved binary positions are measured with respect to the photocenter. In our case, we assume the companion is dark, so the photocenter coincides with the luminous star. This implies that the observed astrometric motion traces the orbit of the luminous visible star around the barycenter, with $a_\text{vis}$ corresponding to its semi-major axis \citep{Platais2024}. This {\em a priori} assumption that the companion is dark may not hold true. However, if the companion is indeed luminous, the true mass of the object will be higher than than what we infer \citep[Equation 4]{Platais2024}.   
For neutron stars and black holes in wide binaries, the expectation is that they make no appreciable contribution to the photocenter position. 

 As this analysis is carried out within a Bayesian framework, we introduce a log-prior, $\ln(prior)$, which encodes our assumptions and constraints on the model parameters. Our prior choices are shown in Table \ref{tab:priors}, where we use uniform priors across most quantities to ensure our model is primarily data driven. The exception is for the proper motions, where we also place an additional constraint based on the velocity dispersion of black hole binaries from the simulations (Section~\ref{sec:simulation}) that each component follows a normal distribution with a one-dimensional velocity of 12.6 km s$^{-1}$, translating to 0.484 mas/yr  centered around the mean motion of the cluster.  
 The sensitivity of our results to this prior are discussed in Section \ref{sec:fit convergence}.
 
We used the \texttt{emcee} package \citep{Foreman-Mackey2013} to sample the posterior probability distribution.
We run the sampler with 350 chains over 150,000 iterations and use a combination of moves: 80\% of the moves done with the ``stretch move" ensemble method and 20\% done with the differential evolution Snooker move. This combination ensures that we sufficiently explore the whole parameter space and avoid becoming stuck in local minima. The effectiveness of this choice is reflected in our Effective Sample Size (ESS), which for each parameter is $\gtrsim$ 1000, which indicates we have a sufficient amount of independent samples and our chains are well mixed.  A more in depth discussion of our orbital fitting methods will be presented in Whitaker et al. ({\em in prep}).

\begin{table}
\centering
\caption{Priors for the orbital parameters of oMEGACat BH-2. An additional normal prior is applied to the proper motions with a one dimensional velocity of 0.484 mas/yr, centered around the mean motion of the cluster (see Section \ref{sec:orbital_fit}). For numerical stability, we limit the $e$ to 0.95 and proper motions to be within 2 mas/yr of the catalog value $\mu_\text{cat}$.}
\label{tab:priors}
\begin{tabular*}{\columnwidth}{ll}
\toprule
Parameter & Prior \\
\midrule
Eccentricity, $e$ & $\mathcal{U}(0, 0.95)$ \\
Inclination, $\cos i$ & $\mathcal{U}(0, 1)$ \\
Semi-major axis, $a_\text{vis}$ (AU) & $\mathcal{U}(1, 100)$ \\
Period, $P$ (yr) & $\mathcal{U}(0.5, 200)$ \\
Longitude of ascending node, $\Omega$ & $\mathcal{U}(0, 2\pi)$ \\
Argument of periastron, $\omega$ & $\mathcal{U}(0, 2\pi)$ \\
Time of periastron passage, $T_0$ & $\mathcal{U}(-\frac{P}{2}, \frac{P}{2})$ \\
Astrometric offsets, $\Delta \text{RA}$, $\Delta \text{Dec}$ (AU) & $\mathcal{U}(-50, 50)$ \\
Proper motions, $\mu_\text{RA}$, $\mu_\text{Dec}$ (mas/yr) & $\mathcal{U}(-2, 2) + \mu_\text{cat}$\\
 & \& $\mathcal{N}(0, 0.484$)\\

\bottomrule
\end{tabular*}

\end{table}

\subsection{The Best-Fit Model} \label{sec:bestfit}

\begin{figure*}[ht!]
    \centering
    \includegraphics[width=1\linewidth]{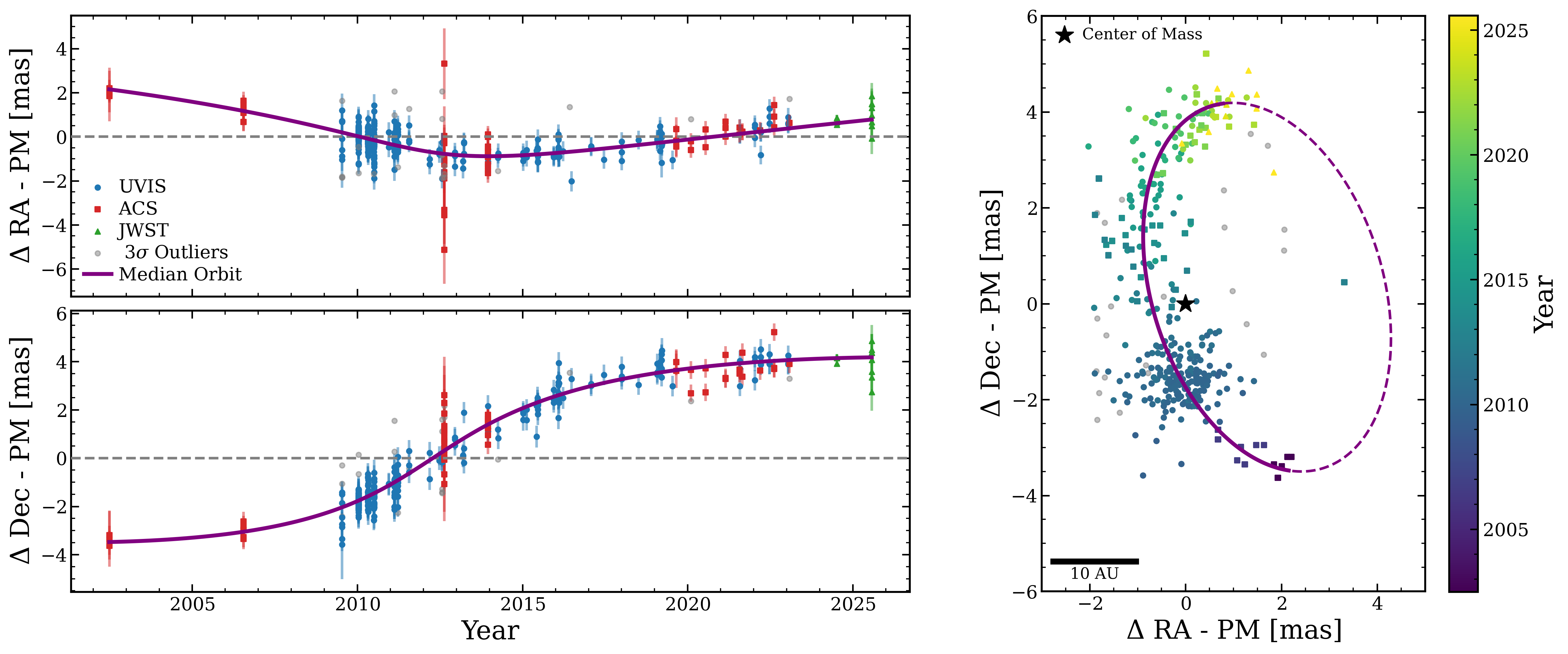}
    
    \caption{Best fit orbital model. \textit{Left} - Our median model (purple line) shown compared to the data after subtracting out the derived proper motion. The marker shape and color corresponds to the instrument used for measurement, and the gray points represent $3 \sigma$ outliers from the median fit. \textit{Right} - Our median orbital model plotted with the on-sky data in the proper motion subtracted frame colored by observation date; the orbit is shown as a dashed line outside our observation window.}
    \label{fig:orbital_fit}
\end{figure*}

\begin{figure*}[ht!]
    \includegraphics[width=1\linewidth]{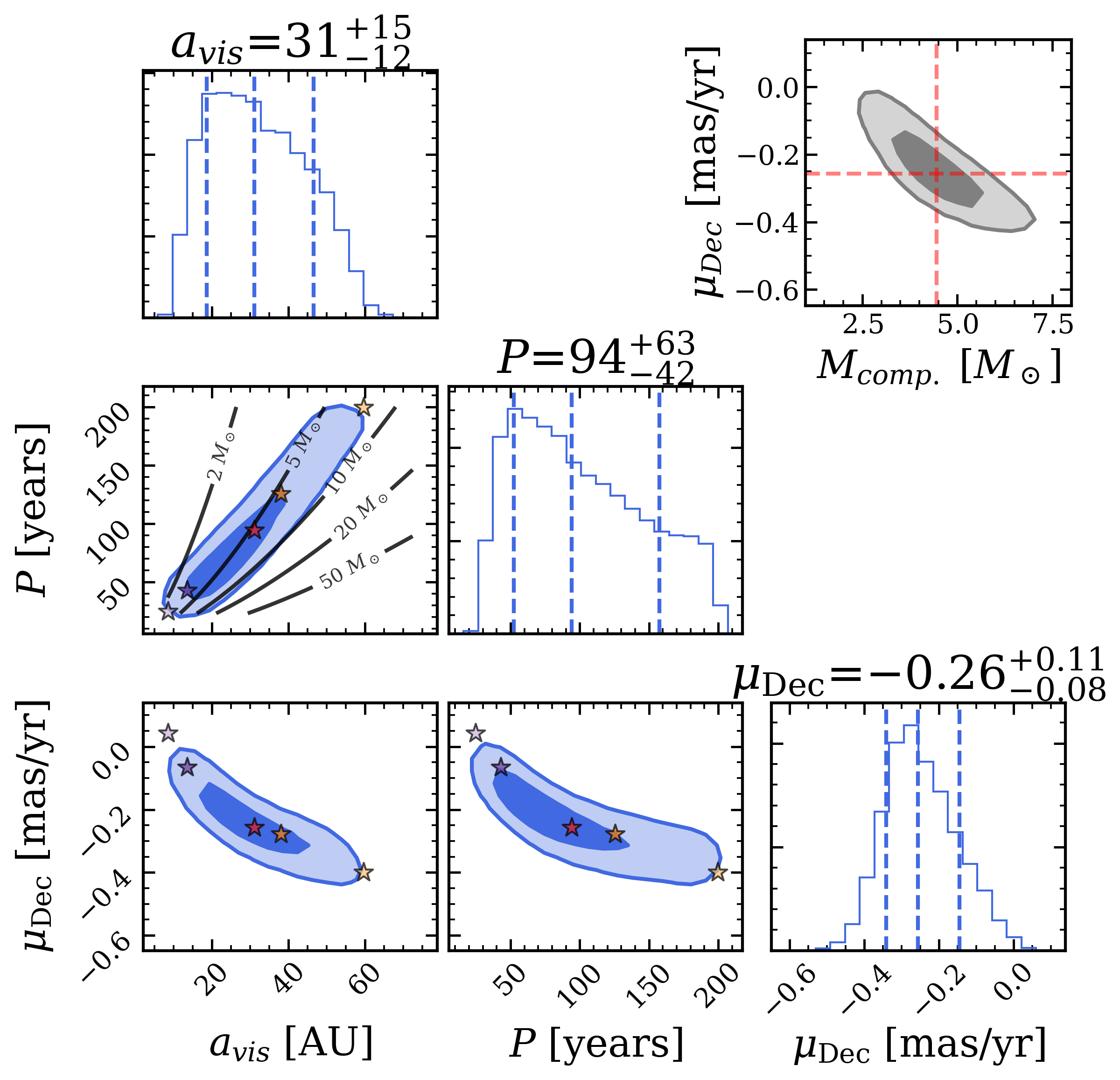}
    \caption{Degeneracies corner plot.  Corner plot showing the degeneracies between Period ($P$), semi-major axis ($a_\text{vis}$), and the declination proper motion. The cutout in the top right shows the degeneracy between the companion mass and the declination proper motion ($\mu_\text{Dec}$). The contours on the $a_\text{vis}$ versus $P$ subplot represent a constant mass corresponding to $M_\text{comp}$, showing the spine of the ellipse closely follows along the $5~M_\odot$ contour. The stars in each panel represent the orbital solutions corresponding to the median, $\pm$1$\sigma$ and $\pm$2$\sigma$ of the joint 2D probability distribution of $a_\text{vis}$ and $P$.}
    \label{fig:corner_plot}
\end{figure*}

Table \ref{tab:paramvals} presents the best-fit solution for our 11 orbital parameters. The reported values represent the medians of the marginalized posterior distribution. The Table also presents 1 and 2$\sigma$ uncertainties for each parameter.  
Our best fit yields a long-period binary system with $P$= $\Period^{+63}_{-42}$ years, a semi major axis of $a_\text{vis}$ = $31^{+15}_{-12}$ AU, an eccentricity of $e$=$\e^{+0.08}_{-0.13}$, and implies a mass of the dark component of $\mcomp^{+1.22}_{-1.01}$~M$_\odot$; we examine the full range of our inferred masses more in Section~\ref{sec:mass_constraints}.

The long orbital period and high eccentricity imply that the visible companion spends most of its orbit at large separations from the dark companion. This is consistent with the smaller astrometric changes found at later epochs as seen in Fig. \ref{fig:orbital_fit}, which shows the median parameter orbital fit compared to the data. As we discuss below in Section \ref{sec:bh_pop_completeness}, this binary was detected at high significance only because the HST data happened to catch the periapsis of the visible star. Despite significant uncertainty on the period, the time of periastron passage is very well constrained: $T_0$ = $\Tperi^{+0.32}_{-0.36}$. The small fractional uncertainty in $T_0$ suggests that the most dramatic orbital motion is observed over our baseline. We empirically calculated the on-sky velocity and acceleration over time and find both peak at the inferred periastron passage time.

Our best fits find a significant offset in the declination proper motion relative to the catalog proper motion, because the catalog fit captured a combination of binary and systemic motion. Hence the inferred declination proper motion is covariant with the semi-major axis and period; this covariance is shown in Fig.~\ref{fig:corner_plot}.  This parameter degeneracy is examined further below in Section~\ref{sec:fit convergence}.

The orbital fit with the median parameters (Fig. \ref{fig:orbital_fit}) result in a reduced chi-squared of $\chi_{\nu}^2 \sim 1.65$. Examining the residuals, we find that our $\chi_{\nu}^2$ value is inflated by the presence of outliers (3-5$\sigma$). However, the bulk of our data are fully consistent with the orbital fit: the 16th/84th percentiles of the error normalized residuals (i.e. ${(x_{obs} - x_{model}})/ {\sigma}$) are $[-0.85,0.97]$ and $[-0.93,0.77]$ in the RA and Dec directions respectively, consistent with a good fit to the data.
 
 We also check the fit's robustness by performing an alternate fit using the nested sampling technique. Nested sampling is effective at discovering multi-modal solutions \citep{Buchner2023}. We use \texttt{UltraNest} \citep{Buchner2021} with the same priors as the MCMC fit (Table \ref{tab:priors}). However, in order to reduce the search space and allow \texttt{UltraNest} to converge, we use only eccentricity, period, and time of periastron as our primary parameters. The remaining parameters are all linear, and can be solved using ordinary least-squares. We propagated the uncertainties of the linear parameters from the posterior samples of the three directly constrained parameters. Despite the very different method, the results of this fit were consistent with the results from MCMC and show no multi-modalities. Hence, we present only the MCMC results here.

\subsubsection{Examining the Degeneracy in Orbital Parameters}
\label{sec:fit convergence}

Our best fit is a long-period system with our baseline covering less than 40\% of the median inferred period---a regime that can lead to major complications in binary characterization \citep{Lucy2014}. However, because our data capture the periastron passage around early 2012, the observations provide a significantly higher dynamical signal than observations at other orbital phases, thus allowing us to constrain the binary parameters more effectively.

The primary degeneracy in our best fit solution is between $\mu_\text{Dec}$, $a_\text{vis}$ and $P$. This degeneracy arises because if the visible star has a high proper motion, especially in declination ($\mu_\text{Dec}$), the data can be well fit by a larger orbit (with both increased $a_\text{vis}$ and a longer period). For such fits, the observed data represent a smaller fraction of the orbit.
Figure \ref{fig:corner_plot} demonstrates this degeneracy in the parameter fits. We note that the velocity dispersion prior on the proper motion has a mild effect on the posterior, disfavoring the highest proper motion (and hence longest period) solutions. Omitting this prior results in only a slightly higher semi-major axis and period, with the inferred mass rising by $\sim 0.1$ $M_\odot$.

To further explore this degeneracy, we select the orbital fits corresponding to the median, $\pm$1$\sigma$ and $\pm$2$\sigma$ values of the joint 2D probability distribution of $a_\text{vis}$ and $P$ along the line of maximum degeneracy (see stars in  Figure \ref{fig:corner_plot}). Selected parameters for these fits are presented in Table \ref{tab:ridge_points}.
The masses found in the table for the $\pm 1 \sigma$ and $\pm 2 \sigma$ values are calculated directly from the corresponding sample. The most compact orbits correspond to periods of about 25 years and modest eccentricity ($e\sim0.3$), while the most extended orbits reflect 200 yr periods with high eccentricity ($e\sim0.8$). \added{Increasing the maximum value allowed by the orbital period prior continues the trend shown in Table \ref{tab:ridge_points}, allowing more long-period solutions with higher proper motions, eccentricities, and orbital periods that lead to larger inferred masses. However, for extremely long period solutions, it becomes increasingly unlikely that we captured periastron passage in our dataset (see Section \ref{sec:bh_pop_completeness}).}

\begin{table}[t]
\centering
\caption{Orbital parameters of oMEGACat BH-2.  Last column includes a comparison to the fit by \citet{Platais2024} (their HSToC-4). Uncertainties given by Platais et al. are formal fitting errors only, whereas we report the 16th and 84th percentiles from the sampled posterior distribution. Priors for the fit parameters are given in Table \ref{tab:priors}.}
\label{tab:paramvals}
\setlength{\tabcolsep}{0pt}
\begin{tabular*}{\columnwidth}{l l l l l l l}
\toprule
Parameter & This\hspace{0.1in}& $-2\sigma$\hspace{0.15in} & $-1\sigma$\hspace{0.15in} & $+1\sigma$\hspace{0.15in} & $+2\sigma$\hspace{0.15in} & P24 \\
          &    Work\hspace{0.1in}       &            &            &            &            & \\
\midrule
$e$                             & $\e$        & 0.45  & 0.58  & 0.80   & 0.83   & 0.3$\pm$0.2 \\
$\cos i$                        & $\cosi$     & 0.36  & 0.39  & 0.47   & 0.53   & — \\
$a_\text{vis}$ [AU]             & $\avis$     & 12.7  & 18.7  & 46.6   & 56.4  & 4.7$\pm$0.4 \\
$P$ [yrs]                       & $\Period$   & 34 & 52 & 157 & 192 & $13 \pm 2$ \\
$\Omega$                        & $\longasc$  & 2.75  & 2.81  & 2.92   & 2.97   & — \\
$\omega$                        & $\argperi$  & 1.44  & 1.53  & 1.67   & 1.75   & — \\
$T_0$ [yr]                      & $\Tperi$    & 2011.4 & 2011.8 & 2012.5 & 2012.8 & — \\
$\Delta RA$ [mas]               & $\deltaRA$  & 0.00  & 0.11  & 0.38   & 0.54   & — \\
$\Delta Dec$ [mas]              & $\deltaDec$ & 0.14  & 0.30  & 0.61   & 0.75   & — \\
$\mu_\text{RA}$ [mas/yr]\hspace{0.05in}        & \deltapmra & -0.07 & -0.03 & 0.02   & 0.04   & — \\
$\mu_\text{Dec}$ [mas/yr]       & -0.26 & -0.41 & -0.34 & -0.15  & -0.04  & — \\
$M_\text{vis}$ [$M_{\odot}$]    & $\mvis$     & —     & —     & —      & —      & 0.78$\pm$0.01 \\
$M_\text{comp}$ [$M_{\odot}$]   & $\mcomp$    & 2.69  & 3.44  & 5.68   & 7.26   & $1.36^{+0.36}_{-0.26}$ \\
\bottomrule
\end{tabular*}
\end{table}

\subsubsection{Mass Constraints} \label{sec:mass_constraints}

\begin{figure}[t!]
    \centering
    \includegraphics[width=\linewidth]{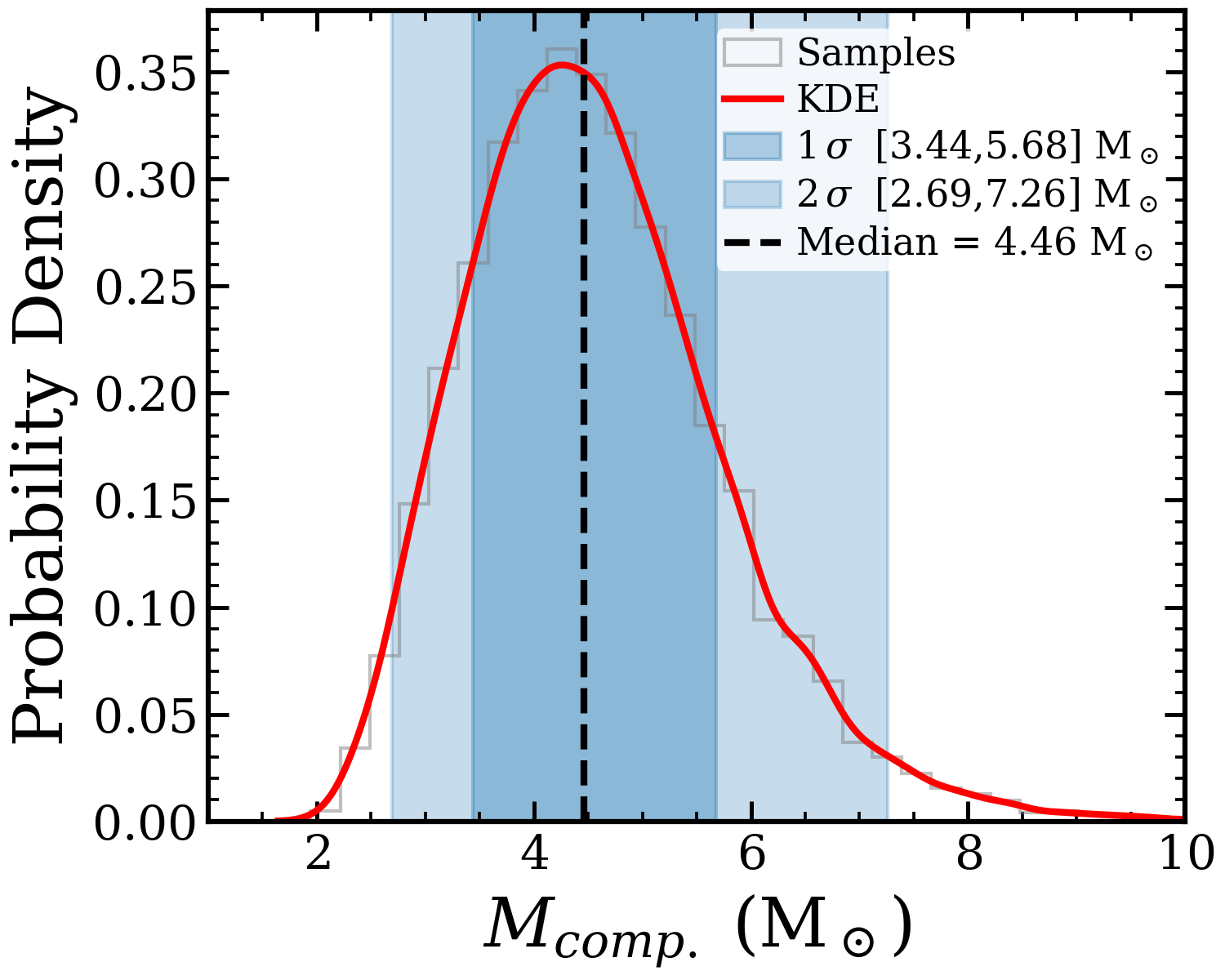}
    \caption{Companion mass.  A histogram showing the posterior distribution of the unseen companion mass and its kernel density estimate. These mass samples are derived from samples drawn from from the $a_\text{vis}$ and $P$ joint posterior distribution.}
    \label{fig:mass_histogram}
\end{figure}

The masses of the binary components are not directly included in the above MCMC fits, but must be separately inferred from the posterior samples of $P$ and $a_\text{vis}$, which determine the total mass of the binary. For each sample, the mass of the unseen binary companion is uniquely determined for a given assumed mass of the observed main sequence star. For our best-fit model, we assume a mass of $0.78 M_{\odot}$ as derived in Section \ref{sec:star_est}.  Fig.~\ref{fig:mass_histogram} shows the inferred companion masses $M_\text{comp}$ from our orbital fits; the best fit median mass is 4.46~M$_\odot$, with a 1$\sigma$ range of 3.44--5.68~M$_\odot$, and a 2$\sigma$ range of 2.69--7.26~M$_\odot$. 

These mass ranges indicate that the invisible companion is almost certain to be a black hole. The highest mass neutron star with a precisely measured mass has $2.08\pm0.07 M_{\odot}$ \citep{Fonseca2021}; slightly higher masses are claimed for some ``spider" neutron star binaries \citep{Romani2026}, but these suffer from modeling challenges and are less certain. Hence there is potentially a tiny sliver of parameter space for which an unusually massive neutron star cannot be excluded with 100\% certainty with our existing data; such solutions would also correspond to the shortest orbital periods, so can be ruled out with only a couple more years of JWST astrometry (see Section \ref{sec:improved_mass_constraints}). 

\added{We also cannot fully rule out scenarios where the invisible object is itself a rare compact object binary containing two neutron stars, a neutron star and a massive white dwarf, or two massive white dwarfs. Several compact binaries with a high total mass are known in globular clusters that contain a pulsar and a compact object companion, such as M15C (total mass $2.7 M_{\odot}$; \citealt{Jacoby2006}), Ter5ao (total mass $3.2 M_{\odot}$; \citealt{Padmanabh2024}), and NGC 1851E (total mass $3.9 M_{\odot}$; \citealt{Barr2024}); in the latter case the invisible object could be either a massive neutron star or a stellar-mass black hole. Any massive neutron star is likely a millisecond pulsar with its mass increased by the recycling process (e.g., \citealt{Tauris2011}), so this scenario is also disfavored by the lack of a pulsar detection at this location \citep{Bernadich2026}. Additional evidence against a millisecond pulsar is the lack of an X-ray source (Section~\ref{sec:X-ray}), since millisecond pulsars are expected to be X-ray sources detectable in deep data, as is true for most of the known $\omega$ Cen pulsars \citep{Zhao2023}. 

Potential contaminating compact binaries are also rare in the \citet{GonzalezPrieto2025} simulation snapshot, with zero such systems at any radii $> 2.7 M_{\odot}$ and only five $> 2.5 M_{\odot}$ (four of these are white dwarf--white dwarf binaries; one is a neutron star--neutron star binary). Hence the predicted abundance of potential contaminating compact binaries is $\sim 0.2\%$ that of black holes. While the prescriptions for neutron star formation and mass accretion used in these simulations are likely imperfect, the observed scarcity of compact binaries with high total mass in other clusters suggests such binaries are truly uncommon.}

For the remainder of the paper we assume the very likely black hole identity of the companion, referring to it as  oMEGACat-BH-2, as this is the second black hole in $\omega$~Cen discovered in the oMEGACat project data, after the central intermediate-mass black hole \citep{Haberle2024IMBH}.

As noted in Section~\ref{sec:star_est}, there is an uncertainty in the mass of the visible star due to the possibility it may be helium enriched, even though the data disfavor this. To quantify the impact of this potential systematic, we repeat our fit using the helium-rich visible star mass estimate of 0.60~M$_\odot$. This reduces the mass of $M_\text{comp}$ by about 6\%, giving a median value of 4.19~M$_\odot$, and a 2$\sigma$ range of 2.50-7.00~M$_\odot$.  Thus even with this relatively large difference in the mass of the visible star, there is not a significant effect on the inferred companion mass or any qualitative change in our conclusions. Any systematic uncertainties due to stellar properties apart from the helium abundance will have an even smaller impact on the inferred companion mass.

\begin{table}[!ht]
\small
\centering
\caption{Representative Orbits Along the $a_\text{vis}$ vs.~$P$ ellipse}
\label{tab:ridge_points}
\setlength{\tabcolsep}{0.75pt} 
\begin{tabular}{lccccc}
\toprule
Fit & Period [yr] & $a_{vis}$ [AU] & $\mu_{Dec}$ [mas/yr] & $e$ & $M_2$ [$M_\odot$] \\
\midrule
$-2\sigma$ & 24   & 8.57  & $0.04$ & 0.31 & 2.01 \\
$-1\sigma$ & 42   & 13.5  & $-0.07$ & 0.56 & 2.37 \\
Median     & $\Period$ & $\avis$ & $-0.26$ & $\e$ & $\mcomp$ \\
$+1\sigma$ & 125  & 38.1  & $-0.28$ & 0.79 & 4.73 \\
$+2\sigma$ & 199  & 59.6 & $-0.40$ & 0.81 & 6.63 \\
\bottomrule
\end{tabular}
\end{table}

\subsection{Comparison With \citet{Platais2024}} \label{sec:platais_comparison}

This system was identified previously by \citet{Platais2024} as a candidate neutron star--main sequence star binary. A comparison between their orbital parameters and the orbital parameters derived in this work is shown in Table \ref{tab:paramvals}. The masses of the visible components are not significantly different, but the other values show strong inconsistencies.

The primary difference in these estimates comes from our more complete dataset, spanning from 2002 to 2025. The HST data we used includes both ACS/WFC and UVIS/WFC3 images in all available filters plus the new JWST NIRCam images. \citet{Platais2024} include only UVIS/WFC3 F606W images taken from 2009 July 15 through 2022 January 11. Notably, the earliest epochs in our data -- taken before the installation of WFC3 in May 2009 -- provide the strongest constraints on the orbital parameters and require a longer orbital period than the 13~yr period found by \citet{Platais2024}. Hence the new data allow us to fully exclude the \citet{Platais2024} solution, favoring a longer period and higher mass companion.

Another difference between our studies is the method used for orbital fitting. \citet{Platais2024} use an unspecified maximum-likelihood algorithm to fit their combined orbital and proper motion model. They note that the algorithm tends to converge to the shortest period compatible with the data and caution that long-period solutions could still be correct. In contrast, the MCMC method we used allows us to explore the full parameter space by sampling the posterior distribution. 

Although the astrometric data from \citet{Platais2024} is not available, we tested using our MCMC method on our UVIS/WFC3 F606W data within the temporal baseline of their observations. The inferred mass of the companion was less well determined in this fit, and is consistent with both the result we present here and the neutron star mass presented by \citet{Platais2024}.

\subsection{Black Hole Population Completeness}
\label{sec:bh_pop_completeness}

As described in Section \ref{sec:bestfit}, we observed oMEGACat BH-2 as it underwent periastron passage, which was likely critical in its discovery. To understand how this may impact our ability to detect other black hole binaries, we used synthetic astrometric observations of simulated binaries.

To create synthetic observations, we first subtract the proper motion and median model from the astrometric data of our visible star to get a realistic observational cadence and characteristic noise. We then inject the orbital signal that arises from the sampled set of orbital parameters. To determine whether this orbital signal would be detectable, we first re-fit the proper motion, using the same method as \citet{Haberle2024}, and then we apply the same detection method described in our upcoming binary search paper (Whitaker, et al. {\em in prep.}). In short, we use a Lomb-Scargle periodogram on the proper motion residuals of the F606W UVIS astrometric data. From the periodogram, we compute a false alarm probability. We consider any system with a false alarm probability $< 10^{-10}$ to be a detection. For context, the false alarm probability of oMEGACat BH-2 is $10^{-28}$, and our binary search finds no other strong black hole binary candidates (Whitaker et al., {\em in prep}).  

\begin{figure}[t!]
    \includegraphics[width=0.45\textwidth]{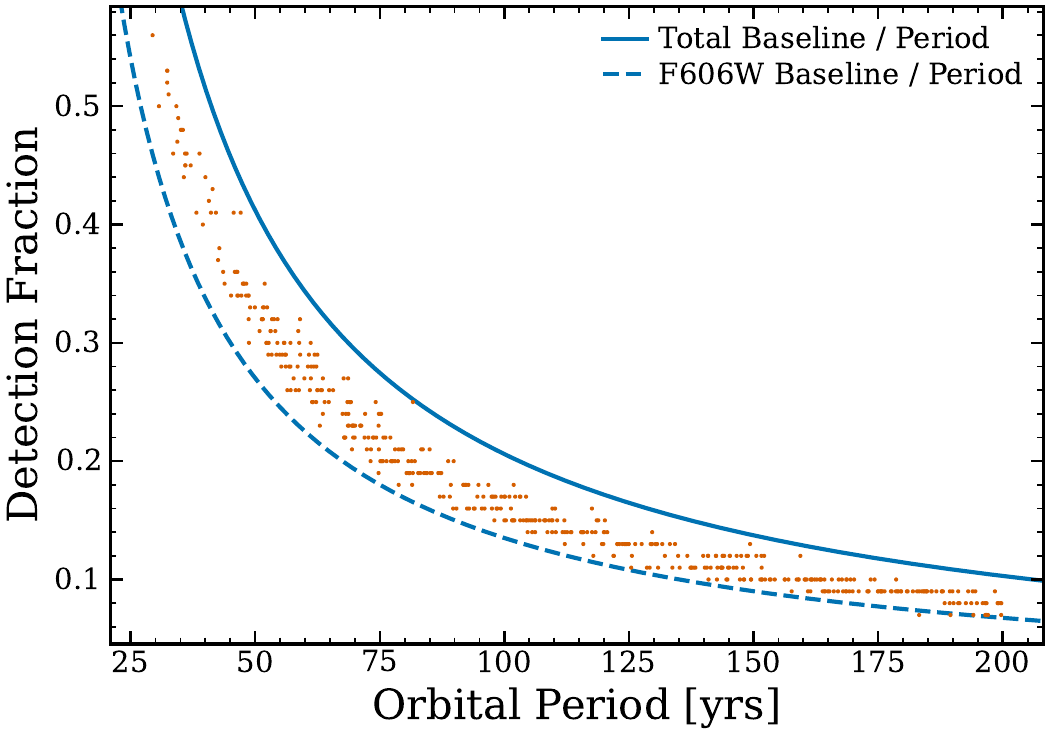}
    \caption{Detection Fraction. The fraction of orbits similar to our posterior samples detected with varying times of periastron passage (see Section \ref{sec:bh_pop_completeness}). For comparison, we show the fraction of the total orbital period covered by the temporal baseline of all our HST observations (20.6 yrs) and the F606W UVIS observations only (13.5 yrs); these were the primary data used to do initial detections.}
    \label{fig:period_completeness}
\end{figure}

We first perform this experiment for the oMEGACat BH-2 system using the posterior samples from our orbital fits but varying the time of periastron passage uniformly over the orbital period. The results are shown in Fig.~\ref{fig:period_completeness}. For orbits near our median period of 94 years, we find we detect the system 15-20\% of the time; this percentage rises towards shorter periods and falls towards longer periods. In essence, we only detect the binary if its periastron is within the dataset and thus our probability of detecting the binary is very similar to the fractional coverage of the orbit. This would tend to favor one of two possibilities: (i) the binary has an orbital period at the shorter end of the allowed range, or (ii)  multiple long-period black hole binaries are present in the cluster. In either of these cases, detecting a single periastron passage in our dataset becomes more likely.

To get a first sense for the sensitivity of our observations to other binary black holes, we used the same data residuals (thus assuming the same visible star in the binary), but then evaluated the detectability of systems across a grid of orbital periods and companion masses between 2 and 40~$M_\odot$. We randomly sampled the remaining orbital parameters. Eccentricities were drawn from the eccentricities of the simulated black hole--star binaries described in Section \ref{sec:simulation}, and the remaining parameters are sampled uniformly in the ranges described by our orbital fit priors in Table \ref{tab:priors}. We repeated the process 1000 times per grid point.

\begin{figure*}[t!]
    \includegraphics[width=1.0\textwidth]{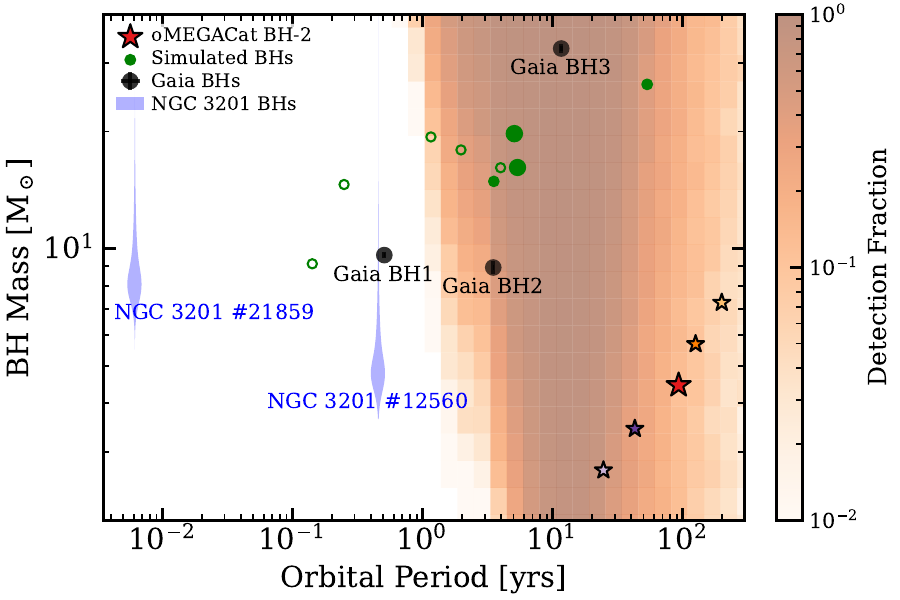}
    \caption{Black Hole Binary Summary.  A comparison of dynamically detected black hole--star binaries and oMEGACat BH-2.  For oMEGACat BH-2, besides the median fit (red star), we show the representative orbits from Table \ref{tab:ridge_points}, using the same colors as Figure \ref{fig:corner_plot}.  For the other observed systems, spectroscopically-detected binaries in the globular cluster NGC 3201 \citep{Giesers2018,Giesers2019} are shown as violin plots to represent the mass-inclination degeneracy, and the three Gaia BHs are also shown \citep{ElBadry2023a,ElBadry2023b,GaiaCollaboration2024}.
    The detection fraction shown in the background assumes a binary with the same secondary mass and observational cadence and uncertainties as for oMEGACat BH-2 and is described in Section~\ref{sec:bh_pop_completeness}.  We also add simulated binaries (green circles; calculated in Section \ref{sec:wcen_bh_population}) from the $\omega$~Cen simulations in Section \ref{sec:simulation}. The size of the marker corresponds to its detection fraction, and open circles indicate no detection; exact values are given in Table~\ref{tab:simulation_detection_fraction}. }
    \label{fig:population_completeness}
\end{figure*}

The results are shown as the background of Fig. \ref{fig:population_completeness}, which plots the detection probability as a function of binary period and black hole mass. We have a high completeness for black hole binary systems of all black hole masses for periods between $\sim$5 and 30 years.  At higher black hole masses, the completeness extends to shorter period systems due to the larger amplitude of the orbital signal; Gaia BH3 like objects would be detectable down to $\sim$1 year periods. As found above, the completeness becomes low at long orbital periods, since coverage of periapsis is needed to securely detect the binary.

We note that these completeness estimates are based on residuals for the visible star in oMEGACat BH-2, which is 
a typical turnoff star, but nonetheless is brighter than the average star in $\omega$~Cen. Hence these results 
do not represent survey-wide completeness. Nonetheless, this test provides us with a useful understanding of our sensitivity to black hole binaries. We discuss this further in Section~\ref{sec:wcen_bh_population}.

\section{Discussion} \label{sec:discussion}

\subsection{oMEGACat BH-2 in Context: Confirmed Stellar-mass Black Holes in Massive Star Clusters}
\label{sec:other_bhs_context}

oMEGACat BH-2 is the first black hole identified astrometrically in a globular cluster, and the longest period black hole binary system known.  
oMEGACat BH-2 is also the first stellar mass black hole identified in $\omega$~Cen, despite the expectation from dynamical models that the cluster contains $\mathcal{O}(10^4)$ black holes within the core of the cluster \citep{Zocchi2019, Baumgardt2019, Banares-Hernandez2025}. 

The only other known stellar-mass black holes in a globular cluster are the two systems discovered and characterized using radial velocities in the cluster NGC~3201 \citep{Giesers2018,Giesers2019}.  
With [Fe/H]=--1.59 \citep{Harris1996}, NGC~3201 has a similar metallicity to the dominant population of stars in $\omega$~Cen. While the masses of these two NGC~3201 black holes are uncertain due to the unknown inclination of their orbits, they both likely have masses below 10~M$_\odot$ (Fig.~\ref{fig:population_completeness}; blue violins). Combined with the $<$7.26~M$_\odot$ (2$\sigma$ upper limit) on oMEGACat BH-2, these masses stand in contrast to expectations of high mass black holes being generated in low-metallicity populations based on stellar evolution models.  For example, \citet{Spera2015} find that for $Z=2 \times 10^{-4}$ to $2 \times 10^{-3}$, spanning the median metallicity of both clusters, the peak black hole mass should be in the range $\sim 20$--40 $M_{\odot}$, depending on the prescription used. This contrast can also be seen in Fig.~\ref{fig:population_completeness} in the difference between the high mass of simulated black holes ($\sim 10$--30 $M_{\odot}$) in the $\omega$~Cen-like cluster from \citet{GonzalezPrieto2025} and the low mass of oMEGACat BH-2 and the two NGC~3201 objects. This mismatch is present despite the 
higher completeness for more massive black holes in both radial velocity and astrometric searches \citep[see Fig.~\ref{fig:population_completeness} \&][]{Saracino2025}.

We draw two conclusions from these observations. First, there must be at least one channel that produces lower mass black holes in low-metallicity systems with modest natal kicks. This could be via a low-compactness progenitor that ejects nearly the entire envelope (e.g., \citealt{Boccioli2024}), or via fallback (e.g., \citealt{Ertl2020}). Such black holes may be common rather than rare.

Second, we observe a \emph{lack} of massive black holes in these two typical metal-poor ([Fe/H] $\lesssim -1.5$) globular clusters. Given the large number of massive black hole--black hole mergers observed in gravitational waves \citep{TheLIGOScientificCollaboration2025}, as well as compelling evidence that the 33 $M_{\odot}$ binary
Gaia BH3 \citep{GaiaCollaboration2024} was formed in a very low-metallicity ([Fe/H] $\sim -2.5$) globular cluster \citep{Balbinot2024}, stellar-mass black holes must sometimes form with the predicted high masses at low metallicity. Further, since the most massive black holes in dense clusters are expected to be preferentially ejected through dynamical interactions (e.g., \citealt{Morscher2013}), it is possible that such interactions have substantially depleted the high end of the black hole mass distribution in both $\omega$~Cen and NGC 3201. Nevertheless, simulations intended to reproduce the properties of these clusters \citep{Kremer2018,GonzalezPrieto2025} all have higher black hole masses than observed. This represents preliminary but increasing evidence that revisions \added{may be needed in the initial black hole mass distribution.}

\subsection{The Dynamical Evolution of oMEGACat BH-2}
\label{sec:dynamical-evolution}

In a star cluster, the dynamical evolution of binaries is governed by a balance between their internal binding energy and the kinetic energy in cluster stars. When the former is larger, the binaries are called ``hard", while in the reverse case they are called ``soft". The standard view of binaries in star clusters follows Heggie's Law: that hard binaries become harder, while soft binaries become softer over time, until they eventually are disrupted \citep{Heggie1975}. However, the law itself is silent on the timescale for the evolution, which will depend on the density of stars. \\

\subsubsection{Critical Velocity} To understand the dynamical evolution of our black hole binary, we first calculate the critical velocity ($v_c$) for the binary \citep{Hut1983} in the context of a binary--single encounter

\begin{equation}
v_c^2 = G \frac{M_{comp} M_{vis}}{M_{inc}}\frac{M_{comp}+M_{vis}+M_{inc}}{M_{comp}+M_{vis}}\frac{1}{a}
\end{equation}

where $M_{inc}$ is the mass of the incoming single star. This is the minimum relative velocity between the binary and single star that can unbind the binary. To calculate $v_c$, we use the posterior samples of 
$M_\text{comp}$ and $a$ measured in Section \ref{sec:results} and $M_\text{vis} = 0.78 M_{\odot}$  from Section \ref{sec:starprops}. For $M_{inc}$, \citet{Baumgardt2023} derive the observed single-star mass function for $\omega$ Cen, finding it is well-fit by a two-component power-law with a break at $0.42 M_{\odot}$, $\alpha = -0.20$ below the break, and $\alpha = -1.37$ above the break (for $N \propto m^{\alpha}$). Assuming a minimum mass of $0.08 M_{\odot}$ and a maximum mass of $0.8 M_{\odot}$, this distribution has a mean mass of $0.37 M_{\odot}$. We draw samples from this distribution for $M_{inc}$.

Using these samples, we find $v_c = 16.5^{+7.6}_{-3.9}$ km s$^{-1}$. Given that the three-dimensional velocity dispersion in the core of the cluster is $\sim 35$ km s$^{-1}$ \citep{Haberle2025}, it is immediately clear that the black hole binary is \emph{soft}, and hence is susceptible to disruption.\\

\subsubsection{Simulations} 
\label{sec:fewbody-simulations}
To understand the dynamical evolution of the binary, we conducted a set of binary--single scattering experiments using {\tt Fewbody} \citep{Fregeau2004}. In addition to the parameters already discussed, we determined the initial relative velocity at infinity $v_{\infty}$ by sampling a one-dimensional velocity dispersion of 12.6~km s$^{-1}$ for the black hole binary and 
20~km s$^{-1}$ for the incoming single star, typical for stars at this projected radius \citep{Haberle2025}; each of these was cut off at the observed escape velocity of 62~km s$^{-1}$ \citep{Haberle2024IMBH}. For each binary, the distance of closest approach was drawn from 0 to $5a$, weighted by area to enable a cross-section calculation. To determine the encounter rate, we calculate the number density $n$ of stars within a three-dimensional radius of 0.7 pc (based on the binary's projected radius of 0.5~pc). We find $n = 5 \times 10^{3}$ pc$^{-3}$ using the 12 Gyr snapshot from the \citet{GonzalezPrieto2025} simulations discussed in Section \ref{sec:simulation}; we include normal stars and white dwarfs but exclude black holes for this calculation (see below). The normal star masses are drawn from the observed mass function and the white dwarf masses from the simulation.

While the encounter timescale is short ($\sim 5$ Myr within $2a$), most encounters have a minimal effect on the binary, due to the typical low mass of the perturbing star and the fly-by nature of the encounter. Overall, we found that single-encounter ionization is possible but rare, with an expected ionization timescale of about 4.5 Gyr.

Given how long this single encounter ionization timescale is, we also ran sequential scattering experiments, where the initial binary properties are the same, but the binary separation and eccentricity are updated after each encounter and then used as input for the next encounter. These sequential scatterings should more accurately capture the gradual softening and disruption of the binary due to the summed effect of many encounters. We find that the median (mean) disruption timescale is about 820 Myr (1.2 Gyr). The disruption timescale has a mild dependence on the initial binary properties; the closest (lowest initial $a$) 25\% of binaries have a median disruption timescale of 1.4 Gyr, compared to 480 Myr for the 25\% most separated (highest initial $a$) binaries. Since this is much shorter than the single encounter ionization timescale, gradual softening is clearly the dominant process in eventually disrupting the binary. 
In addition to gradual softening, exchange encounters also occur, but owing to the high masses of the binary components compared to typical cluster stars, they are uncommon.

These scattering experiments did not include neutron stars or stellar-mass black holes as the incoming object. Neutron stars are not numerous enough (only 28 at a three-dimensional radius $< 0.7$~pc) to be relevant. The situation for stellar-mass black holes is less certain. In the \citet{GonzalezPrieto2025} simulation the mass density of stellar-mass objects within the central 0.7 pc is dominated by 
stellar-mass black holes; it is a factor of two higher than normal stars + white dwarfs. As discussed in Section \ref{sec:other_bhs_context}, these black holes are also massive, with a median mass of $18 M_{\odot}$, so can efficiently ionize binaries in encounters. 

We re-ran the scattering experiments assuming an incoming population of black holes with masses and a number density drawn from the \citet{GonzalezPrieto2025} simulation; they are assumed to have an identical velocity dispersion to the target black hole binary. We obtain somewhat different results: many encounters result in an ionization, with a median timescale to ionization of 170 Myr, a factor of five faster than found above for non-black hole encounters. This is mostly determined by the high masses of the black holes: if we instead assume the black holes all have masses like our observed object ($4.5 M_{\odot}$) then the lifetime becomes much longer, about 660 Myr, comparable to the non-black hole disruption timescale.

These exact numbers should not be taken too seriously: the black holes in the simulation are more centrally concentrated than the stars, so the encounter frequency between oMEGACat BH-2 and other black holes will depend crucially on the actual orbit of the binary. In addition, we already have evidence from the low mass of oMEGACat BH-2 that the black hole mass distribution in the simulations is likely inaccurate. The true number of stellar-mass black holes is also uncertain. These simulations do suggest that the survival of soft binaries in the core of $\omega$ Cen can potentially set limits on the population of invisible stellar-mass black holes, but this would require understanding the formation rate of soft binaries as well.

\subsubsection{A Soft Binary With a Long Enough Life} The $\sim 800$ Myr median timescale to disruption is sufficiently short that it is very unlikely the black hole binary could be primordial, but instead was almost certainly formed dynamically. On the other hand, the timescale is sufficiently long that we are not observing it at a particularly privileged time in the cluster: despite the fact that the binary is soft, it can survive long enough to be observed.

\subsection{The Total Black Hole Population in $\omega$ Cen}
\label{sec:wcen_bh_population}

We have already assessed how detectable oMEGACat BH-2 is as a function of orbital period and black hole mass in Section~\ref{sec:bh_pop_completeness}. To further understand what the detection of oMEGACat BH-2 means for the total black hole population in {\wcen}, we create synthetic astrometric and spectroscopic observations of each black hole from the simulation of \citet{GonzalezPrieto2025} to determine a detection probability for each system. These simulated black hole binaries differ in several ways from oMEGACat BH-2: in addition to having more massive black holes, they have shorter orbital periods, as well as a wide range of visible star masses.

We create synthetic astrometric observations of these binaries using the same process described in Section \ref{sec:bh_pop_completeness}. However, unlike those simulations, here
we use real observations of randomly selected stars with a similar mass to the secondary stars in the simulation, after checking that any real stars with potential orbital signals are excluded. This ensures we have a realistic noise profile and observational cadence for stars of different masses in the simulation. We note that because all the simulated black holes are found in a radial range with relatively uniform astrometric coverage, we do not match the data to simulations explicitly in distance from the cluster center. The component masses, semi-major axis, orbital period, and eccentricity are specified for each system from the simulations, but the other orbital parameters are randomly sampled as before to get the simulated orbital signal.

\begin{table}
\centering
\caption{Simulated BH Detectability. The visible star mass $M_{\rm vis}$, companion mass $M_{\rm comp}$, orbital period $P$, semi-major axis $a$, eccentricity $e$, on-sky distance from the center $R_{proj}$, astrometric detection fraction $f_{\rm astro}$, and spectroscopic detection fraction $f_{\rm RV}$ are shown for each of the BH-star binaries (Section \ref{sec:wcen_bh_population}). Values are shown only for the last snapshot of the simulation.}
\begin{tabular}{cccccccc}
\hline
$M_{\rm vis}$ & $M_{\rm comp}$ & $P$ & $a$ & $e$ & $R_{proj}$ & $f_{\rm astro}$ & $f_{\rm RV}$ \\
$\mathrm{M_{\odot}}$ & $\mathrm{M_{\odot}}$ & $\mathrm{yr}$ & $\mathrm{AU}$ &  & $\mathrm{pc}$ &  &  \\
\hline
\hline
0.249 & 17.94 & 1.97  & 4.14  & 0.88 & 0.036 & 0.0   & 0.0 \\
0.308 & 26.45 & 53.3  & 42.4 & 0.97 & 0.351 & 0.041 & 0.0   \\
0.516 & 16.16 & 5.37  & 7.84  & 0.59 & 1.009 & 0.818 & 0.271 \\
0.127 & 16.14 & 3.98  & 6.37  & 0.95 & 0.748 & 0.0   & 0.0   \\
0.566 & 19.75 & 5.08  & 8.06  & 0.76 & 2.126 & 0.847 & 0.047 \\
0.368 & 9.138 & 0.142 & 0.577 & 0.85 & 0.596 & 0.0   & 0.0 \\
0.429 & 14.89 & 3.53  & 5.76   & 0.82 & 1.428 & 0.014 & 0.0 \\
0.468 & 19.37 & 1.16  & 2.99  & 0.44 & 2.268 & 0.0   & 0.0 \\
0.593 & 14.61 & 0.249 & 0.981 & 0.69 & 2.459 & 0.0   & 0.016 \\
\hline
\end{tabular}
\label{tab:simulation_detection_fraction}
\end{table}

We repeat this 1000 times and measure the detection fraction for each simulated binary using the same log(FAP) $< -10$ threshold discussed in Section~\ref{sec:bh_pop_completeness}. These fractions are listed in Table \ref{tab:simulation_detection_fraction} and shown in Fig. \ref{fig:population_completeness} as the marker size for each simulated black hole. Only four of the nine systems in the simulated snapshot have a non-zero detection fraction. Two of the systems are detected $\sim80\%$ of the time, and the other two are detected $<5\%$ of the time. Three undetected systems have an orbital period $<1.5$ yrs, so the orbital signal is small, and the other two have visible star masses $<0.25\,M_\odot$, so the star is faint, causing noisier astrometry. To improve the statistics, we repeated this process using all 50 simulation snapshots, yielding 474 total black hole--visible star binaries. Overall, we find that the expected number of black hole--visible star binaries detectable in our astrometric dataset is $1.5\pm0.4$; the listed uncertainty is the scatter among epochs.

\citet{Wragg2024} found 275 objects with variable radial velocities in {\wcen} that were likely binaries. \citet{Saracino2025}
were able to characterize 19 of these, finding the rest could not be characterized (i.e., 
the binary parameters could not be determined using the available data). No black hole candidates were identified in these data. To interpret these radial velocity non-detections, we performed a parallel analysis to that done for the astrometry, injecting the simulated orbital signal from the simulation into the \citet{Saracino2025} radial velocities for {\wcen}, and determining detectability using the criteria described by \citet{Wragg2024}. Unlike the case for the astrometry, since the radial velocity data have a limited radial extent that does not fully cover the region where the black hole--star binaries are found in the simulations, an areal completeness correction is needed. This was computed by taking 100 radial bins and counting what fraction of 1000 uniformly sampled points in each bin fall within the radial velocity observational footprint. After applying this correction, we find that the expected number of spectroscopically detected black hole--star systems is $0.4\pm0.1$. The corrected detection fractions for each simulated black hole--star system in the final snapshot and the simulated on-sky radius of each system are shown in Table \ref{tab:simulation_detection_fraction}. Only one of the systems with no astrometric detectability is detectable via radial velocities, while two of the four systems detected via astrometry do have some velocity detection probability. 

To summarize, based on the simulation predictions, the expected number of black hole--star binaries detectable in this astrometric survey is $1.5\pm0.4$ (compared to 1 observed), while in radial velocities the predicted population is $0.4\pm0.1$ (compared to zero observed). Thus the detection of one astrometric black hole in the existing astrometric and radial velocity of $\omega$~Cen is consistent with the number of black hole binaries in the \citet{GonzalezPrieto2025} simulations.

How to interpret this agreement is unclear: there is already a known mismatch between the non-black hole binary period distributions of Cluster Monte Carlo simulations relative to the observed period of binaries \citep{Ye2022,Muller-Horn2025, Saracino2025}, with the simulations over-predicting the number of short-period binaries relative to the observations. This could lead to a similar over-prediction of the number of velocity-detectable black hole--star binary systems. In addition, we already noted in Section \ref{sec:other_bhs_context} the mismatch between the mass distribution of the simulated black holes and observations. 

Nonetheless, both the mock analysis of the simulations presented here, as well as the detection analysis presented in Section~\ref{sec:bh_pop_completeness}, show that there is at least a small population of black hole binaries in $\omega$~Cen, of which oMEGACat BH-2 is only the first system to be discovered. This can be readily seen even apart from the uncertainty in the period distribution of binaries: black holes are expected to pair with a range of low-mass secondaries, seen in simulations of both NGC 3201 \citep{Kremer2018} and $\omega$~Cen \citep{GonzalezPrieto2025}. Hence the detection of turnoff stars as secondaries in two of the three of these binaries likely reflects an observational bias. Binaries with less massive secondaries are present, but are harder to discover. 

Taking the larger step of translating this inferred population of black hole--visible star binaries to the dominant population of single black holes is even more challenging, given 
the predicted lack of correlation between black hole binaries and the parent population of black holes \citep{Morscher2015,Chatterjee2017,OConnor2026}. It seems noteworthy that black holes have as of yet only been discovered in the low-density clusters expected to have the largest current black hole populations, which is perhaps in mild tension with the dynamical prediction that the number black hole--visible star binaries should not strongly depend on the number of black holes.

\section{Conclusions and Future Directions} \label{sec:conc}

We have presented the discovery of a low-mass black hole binary in $\omega$ Cen, the first stellar-mass black hole discovered in a globular cluster via astrometry and only the third to be dynamically confirmed in a Galactic globular cluster.

\subsection{Improved Mass Constraints on oMEGACat BH-2}
\label{sec:improved_mass_constraints}
Despite the wealth of data we have on the binary, the lack of a complete orbit translates to a degeneracy between proper motion and mass. Here we discuss how other existing and future data can help us further constrain the mass of oMEGACat BH-2. 

Future JWST data will be useful. The existing JWST program (GO-8322) provided a first epoch used here, and this program will obtain two more years of observations, likely in 2027 and 2028. These data will likely already be sufficient to rule out (or favor) the shortest period orbits currently consistent with our data. \added{The shortest periods correspond to the lowest masses, so these JWST data could also exclude much of the parameter space for the unlikely (but not impossible) potential contaminant scenarios involving other compact objects discussed in Section \ref{sec:mass_constraints}.}

 \begin{figure}[ht!]
    \centering
    \includegraphics[width=1\linewidth]{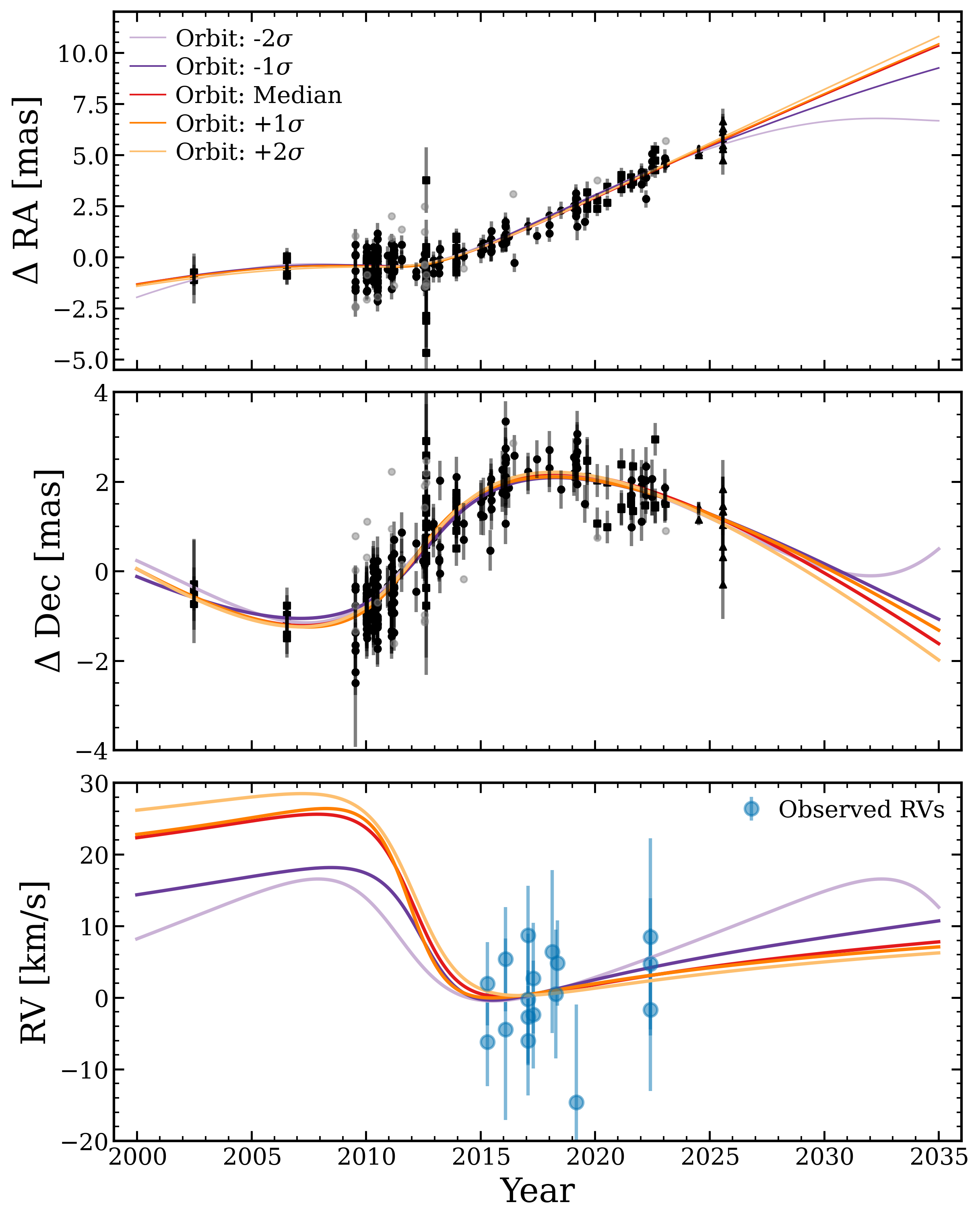}
    \caption{Future model constraints. Orbital fits corresponding to the median and the $\pm$ 1 $\sigma$ and $\pm$ 2$\sigma$ for the joint 2D probability distribution of $a_\text{vis}$ and $P$ as given in Table~\ref{tab:ridge_points}. The models are extended to 2035 to illustrate the utility of future measurements to further constrain the orbit. \textit{Top \& Middle Panels:} These show the relative position over time without removing the proper motion; data points are plotted using the same symbols as Fig.~\ref{fig:orbital_fit}.  \textit{Bottom Panel:} Predicted radial velocity curves for the same 5 orbits. The modeled curves include a fitted offset to the observed velocities (Table~\ref{tab:vrad}).}
\label{fig:orbitalsamples}
\end{figure}

Another potential angle is radial velocities. The bottom panel of Fig.~\ref{fig:orbitalsamples} shows the expected radial velocity from the five orbits (median,$\pm1\sigma$,$\pm2\sigma$) presented in Table~\ref{tab:ridge_points}. Unfortunately, the existing MUSE data are not useful: these data show reduced $\chi^2$ values ranging between 0.43 and 0.45 for the five models plotted. We note a sign degeneracy of the astrometric fit means that the radial velocities can be flipped; we presently choose the family of orbits where the current trend is towards positive velocities. Future, precise radial velocities could determine the correct family of solutions, and could also help constrain the shorter orbital periods.

\subsection{Other Black Holes in $\omega$ Cen and Beyond}

As discussed above, the existing radial velocity and astrometric surveys of $\omega$ Cen are insensitive to the majority of the predicted black hole--visible star binaries.

Owing to the long time baseline of the existing astrometric dataset, it is difficult to improve on its sensitivity to long period orbits anytime soon. JWST offers improved precision, especially of fainter stars, so a JWST astrometric monitoring program with $\gtrsim$10 epochs could enable the detection of black hole binaries with less massive secondaries down to periods of $\sim$100 days. 

Looking to other globular clusters, no comparable HST astrometric datasets exist, but new more sensitive JWST observations spanning a few years would enable detection of black hole binaries with
periods $\mathcal{O}$(1 year). Existing HST astrometry can also play a useful role: we have shown here that the proper motion uncertainty plays a significant role in transforming the observed on-sky accelerations into mass constraints. Therefore, using JWST to target clusters that have long-baseline HST astrometry can also enable identification of longer period black holes like oMEGACat BH-2.
Even without getting new data, existing limited HST astrometric datasets could still show evidence for black hole binaries, especially in lower density clusters where long-period systems may be present. Indeed, that oMEGACat BH-2 was discovered in a soft orbit shows that the simplistic hard/soft boundary should not be taken too literally when considering binary searches.

Other approaches and facilities may also yield additional cluster black hole binary detections.  Gaia DR4 is widely expected to identify many black hole binaries \citep{Nagarajan2025}. However, due to crowding, Gaia typically performs poorly in cluster cores  \citep{Haberle2024}. Given the expectation that cluster black holes will mass segregate to cluster cores, this suggests that Gaia will not be sensitive to most cluster black hole binaries, but this expectation should certainly be tested explicitly.  

Existing radial velocity searches for binaries in globular clusters have been enabled through multi-epoch MUSE measurements in NGC~3201, $\omega$~Cen, and 47 Tuc \citep{Giesers2019, Wragg2024,Muller-Horn2025}, but have only detected the two BHs in NGC~3201.  Many of the detected radial velocity binaries in $\omega$~Cen do not have strong constraints on their orbits, but short period ($<1$ year) black hole binaries should be characterizable with these data as long as the visible companion is a brighter star \citep{Saracino2025}.  Analysis of existing data on additional clusters and additional monitoring could both lead to new black hole binary detections.  

In any case, making further progress on characterizing the population of black holes in globular clusters will require new detections. The identification of oMEGACat BH-2 shows that the age of astrometric black hole discovery in globular clusters has begun.

\begin{acknowledgements}

\added{We acknowledge useful comments from an anonymous referee that helped improve the paper.} Work on this paper by M.W., E.K., A.S., and Z.F. was supported by a Hubble Space Telescope archival grant HST-AR-17033, A.S. also acknowledges support from JWST-GO-08322. J.S. acknowledges support from NASA grant 80NSSC21K0628. S.S. acknowledges funding from the European Union under the grant ERC-2022-AdG, ‘StarDance: the non-canonical evolution of stars in clusters’, Grant
Agreement 101093572, PI: E. Pancino. This research was supported in part by grant NSF PHY-2309135 to the Kavli Institute for Theoretical Physics (KITP); we acknowledge useful conversations at KITP with Kareem El-Badry, Casey Lam, Jeff Andrews, and Jessica Lu. \added {We thank Kareem El-Badry, Jessica Lu, and Johanna M\"{u}ller-Horn for helpful comments on a draft that helped improve the paper.} The support and resources from the Center for High Performance Computing at the University of Utah are gratefully acknowledged.

\end{acknowledgements}

\bibliographystyle{aasjournalv7}
\bibliography{main}{}

\end{document}